\documentclass[11pt]{article}

\PassOptionsToPackage{numbers, compress}{natbib}
\usepackage[margin=1in]{geometry}
\usepackage{natbib}

\usepackage[utf8]{inputenc}
\usepackage[T1]{fontenc}
\usepackage{authblk}
\usepackage[hypertexnames=false]{hyperref}
\usepackage{url}
\usepackage{booktabs}
\usepackage{amsfonts}
\usepackage{nicefrac}
\usepackage{microtype}
\usepackage{xcolor}
\usepackage{amsmath}
\usepackage{amssymb}
\usepackage{multirow}
\usepackage{enumitem}
\usepackage{threeparttable}
\usepackage{graphicx}
\usepackage{bibentry}
\usepackage{listings}
\usepackage{float}
\usepackage{changepage}
\usepackage{tikz}
\usetikzlibrary{arrows.meta, positioning, calc, decorations.pathreplacing, backgrounds, fit}

\title{Simulating Eating Disorder Patients with LLMs: \\
Evaluating Psychological Persona Stability in Multi-Turn Conversations}

\date{Preprint}

\author[1,2]{Jennifer Haase\thanks{\texttt{jennifer.haase@hu-berlin.de}}}
\author[1,3]{Jana Gonnermann-Müller}
\author[4]{See Heng Yim}
\author[3]{Nicolas Leins}
\author[1,2]{Jan Mendling}
\author[3]{Sebastian Pokutta}

\affil[1]{Weizenbaum Institute, Berlin, Germany}
\affil[2]{HU Berlin, Berlin, Germany}
\affil[3]{Zuse Institute Berlin, Berlin, Germany}
\affil[4]{Department of Psychology, University of Hong Kong, Hong Kong}

\begin{document}
\nobibliography*

\maketitle

\begin{abstract}
Large language model (LLM)-based simulations of clinical patients are increasingly used for research and training, yet their validity requires \emph{persona stability}: coherent maintenance of an assigned psychological profile across and within conversations. We evaluate this prerequisite using eating disorder personas grounded in five published case vignettes, a dual-assessment framework (self-report + independent observer ratings), and validated psychometric instruments (EDE-Q) with known ground-truth scores. Across six LLMs and two experiments (between-conversation stability (Exp.~I) and within-conversation stability (Exp.~II)), we find that LLMs are paradoxically \emph{too stable and too inaccurate}: variability is negligible, yet all models systematically overshoot ground-truth severity by 12--30\% of the scale range (0.7--1.8 points on a 0--6 scale). The mechanism is selective stereotyping: models differentiate cases on \emph{behavioural} items (dietary restraint) but maximise \emph{cognitive-affective} items (body dissatisfaction, weight preoccupation) at ceiling regardless of case severity. Additional conversational context does not improve accuracy; it compounds the overshoot. LLMs can portray severe eating pathology but lack a representation of moderate clinical presentations, a ``missing middle''.
\end{abstract}

\section{Introduction}
\label{sec:introduction}

Large language models (LLMs) are increasingly deployed as proxies for human participants, serving as simulated survey respondents \cite{parkGenerativeAgentSimulations2024, sunRandomSiliconSampling2024}, experimental subjects \cite{manningAutomatedSocialScience2024, yeykelisUsingLargeLanguage2024}, social agents \cite{parkGenerativeAgentsInteractive2023}, and simulated patients for clinical training \cite{lozoyaLeveragingLargeLanguage2025, ajluniArtificialIntelligencePsychiatric2025}. These applications rest on a shared but under-examined assumption: that LLMs maintain stable, coherent personas across extended interactions. If an assigned psychological profile drifts mid-conversation or shifts unpredictably across sessions, downstream conclusions about social phenomena, clinical competency, or treatment effects become artifacts of model inconsistency rather than genuine findings \cite{gonnermann-mullerStablePersonasDualAssessment2026}. Persona stability is therefore not merely a desirable property but a \emph{prerequisite} for any serious use of LLMs as behavioural proxies \cite{weidingerEvaluationScienceGenerative2025}.

Early evidence is mixed. Persona benchmarks report stable responses \cite{huangDesigningLLMAgentsPersonalities2024, wangRoleLLMBenchmarkingEliciting2024}, but adding personas does not reliably improve task performance \cite{zhengWhenHelpfulAssistant2024}, formatting changes can swing accuracy by up to 76 points \cite{sclarQuantifyingLanguageModels2024}, and Reinforcement learning from Human Feedback (RLHF) sycophancy compounds across multi-turn dialogues \cite{sharmaUnderstandingSycophancyLanguage2025, liuTRUTHDECAYQuantifying2025}. In clinical settings, \citet{yunPersonaDriftAdaptive2026} documented persona drift in Cognitive Behavioural Therapy chatbots, and \citet{gonnermann-mullerStablePersonasDualAssessment2026} found that ADHD persona self-reports remain stable while observer-rated consistency lags behind.

Current evaluation paradigms cannot detect such failures. Simulated patient systems are evaluated on dialogue realism or diagnostic accuracy \cite{lozoyaLeveragingLargeLanguage2025, ozgunTrustworthyAIPsychotherapy2025, acevedoEvaluatingEfficacyChatGPT352026}, not psychological stability. Persona benchmarks target fictional or personality-trait settings \cite{samuelPersonagymEvaluatingPersona2024, abdulhaiConsistentlySimulatingHuman2025} and use generic user-simulation roles rather than clinically grounded profiles with validated instruments. What is missing is repeated, multi-source psychological measurement of simulated patient behaviour using clinical instruments and stability metrics.

We address this, we work with eating disorders (EDs), psychiatric conditions with the highest mortality rate of any mental illness \cite{fairburnTransdiagnosticTheoryEating2003}, high stigma, and significant training gaps among clinicians. Artificial Intelligence (AI)-simulated ED patients have been shown to reduce trainee anxiety and increase self-efficacy in psychiatric education \cite{ajluniArtificialIntelligencePsychiatric2025}, yet the psychological validity and stability of such personas remain unevaluated. Meanwhile, commercial AI chatbots have been documented reinforcing disordered eating behaviours and failing to capture emotional nuance \cite{deneckeUncoveringAIsHidden2026, winecoffSymptomsSystemsExpertGuided2025}, underscoring the need for rigorous evaluation before deployment.

We ask: \emph{How stably do LLMs maintain assigned eating disorder personas across independent conversations and throughout extended conversations, and which components of a persona description are necessary for clinically plausible simulation?}
We do not study therapy efficacy or multi-agent therapy pipelines; we study whether LLMs can simulate psychologically coherent ED personas, a prerequisite for any downstream application.

Our contributions are: (1)~A \textbf{dual-assessment framework} jointly measuring persona self-reports and independent LLM observer ratings using the EDE-Q, a clinical eating disorder questionnaire,  with multi-provider rater aggregation to address LLM-as-judge bias \cite{zhengJudgingLLMasaJudgeMTBench2023, panicksseryLLMEvaluatorsRecognize2024}. (2)~A \textbf{theoretically grounded prompt richness manipulation} based on Fairburn's transdiagnostic model \cite{fairburnTransdiagnosticTheoryEating2003}, varying persona descriptions from full vignettes to minimal diagnostic parameters, with all assessment scores excluded to prevent circularity. (3)~\textbf{Stability experiments} across six LLMs, five ED personas, and three prompt richness levels, with variance decomposition identifying the relative contribution of model, persona, and prompt to score variation.

\section{Related Work}
\label{sec:related}

\paragraph{Persona simulation and stability.}
Persona assignment meaningfully alters LLM behaviour \cite{salewskiInContextImpersonationReveals2023}, and benchmarks now evaluate fidelity across fictional and personality-trait settings \cite{wangRoleLLMBenchmarkingEliciting2024, wangIncharacterEvaluatingPersonality2024, samuelPersonagymEvaluatingPersona2024, liQuantifyingAIPsychology2024, huangDesigningLLMAgentsPersonalities2024}. However, these predominantly evaluate surface-level consistency. Multi-turn stability remains poorly understood: \citet{abdulhaiConsistentlySimulatingHuman2025} found that line-to-line coherence stays high even as global persona alignment degrades, with mental health personas proving the most challenging domain. \citet{yunPersonaDriftAdaptive2026} documented systematic persona drift in CBT chatbots. Most directly relevant, \citet{gonnermann-mullerStablePersonasDualAssessment2026} proposed dual assessment for ADHD simulations, finding stable self-reports but declining observer ratings, a dissociation motivating our design. We extend this paradigm to eating disorders with validated instruments and a theoretically grounded prompt richness manipulation \cite{fairburnTransdiagnosticTheoryEating2003}.

\paragraph{Simulated patients.}
LLM-based simulated patients are deployed for clinical training \cite{lozoyaLeveragingLargeLanguage2025, mitsopoulosPsychologicallyValidGenerativeAgents2023, paglieriPersonaGeneratorsGenerating2026}, multi-agent diagnosis \cite{ozgunTrustworthyAIPsychotherapy2025, xuLingxiDiagBenchMultiAgentFramework2026}, and even direct-to-patient therapy \cite{acevedoEvaluatingEfficacyChatGPT352026}. A common limitation is that these systems are evaluated on conversation realism, diagnostic accuracy, or symptom reduction, not on the psychological validity or temporal stability of the simulated persona itself. Our work addresses this gap.

\paragraph{LLM-as-judge.}
LLM judges achieve over 80\% agreement with human preferences \cite{zhengJudgingLLMasaJudgeMTBench2023} but exhibit position bias, verbosity bias, and self-enhancement bias; the latter formalised by \citet{panicksseryLLMEvaluatorsRecognize2024}, who showed a linear correlation between self-recognition ability and self-preference magnitude. We mitigate these concerns by using three raters from different providers, aggregating scores, and withholding persona prompts from observers.

\paragraph{Prompt sensitivity.}
Meaning-preserving formatting changes can swing LLM accuracy by up to 76 percentage points \cite{sclarQuantifyingLanguageModels2024}, and selecting \emph{which} persona attributes to include matters as much as format \cite{beckSensitivityPerformanceRobustness2024, luzdearaujoPrincipledPersonasDefining2025}. The wide space of plausible prompting decisions constitutes analytic flexibility analogous to researcher degrees of freedom \cite{cumminsThreatAnalyticFlexibility2026}. These findings motivate repeating each condition across 50 independent runs and systematically varying prompt \emph{content}.

\paragraph{Clinical grounding.}
Persona construction is grounded in Fairburn's transdiagnostic cognitive-behavioural model \cite{fairburnTransdiagnosticTheoryEating2003}, which identifies the \emph{over-evaluation of eating, shape, and weight} as the core psychopathology shared across ED diagnoses. The model distinguishes active maintaining mechanisms (dietary restraint, binge--purge cycles, body checking, mood triggers) from predisposing factors and biographical context, a distinction we operationalise as prompt richness levels (Section~\ref{sec:ablation}).

Five published ED case vignettes serve as personas, spanning Bulimia Nervosa (BN and AN-BP), Binge Eating Disorder (BED/NES and BED), and OSFED/Purging Disorder (PD). All report EDE-Q ground-truth (GT) scores (5.25 for BN to 2.64 for PD; see Table~\ref{tab:scores}). Each persona is structured along nine clinical dimensions (diagnosis, over-evaluation, restraint, binge pattern, compensatory behaviours, body checking, cognitive beliefs, emotional patterns, contextual factors); full vignettes and source citations are in Appendix~\ref{app:vignettes}.

\section{Methodology}
\label{sec:methodology}

\begin{figure}[t]
\centering
\resizebox{\textwidth}{!}{%
\begin{tikzpicture}[
    >=Stealth,
    block/.style={rectangle, draw=gray!70, rounded corners=2pt,
                  align=center, font=\small, fill=#1, line width=0.5pt,
                  inner sep=4pt},
    block/.default={white},
    arrow/.style={->, thick, gray!60}
]

\node[block=blue!6] (vig) at (0, 0) {
    \textbf{5 Vignettes}\\[1pt]
    {\scriptsize BN, AN-BP, BED/NES, BED, PD}
};
\node[block=orange!8] (rich) at (6, 0) {
    \textbf{$\times\,$3 Prompt Richness}\\[1pt]
    {\scriptsize Full $\mid$ Core $\mid$ Minimal}
};
\node[block=green!6] (llm) at (12, 0) {
    \textbf{$\times\,$6 LLMs}\\[1pt]
    {\scriptsize 15 conditions $\times$ 6 models}
};

\draw[arrow] (vig) -- (rich);
\draw[arrow] (rich) -- (llm);

\node[block=purple!6] (exp1) at (5.5, -1.8) {
    \textbf{Exp.\,I} Between-conv.\\[1pt]
    {\scriptsize $N\!=\!50$ independent runs}
};
\node[block=red!6] (dual) at (14, -2.6) {
    \textbf{Dual Assessment}\\[1pt]
    {\scriptsize Self-report + 3 LLM observers}\\[0pt]
    {\scriptsize EDE-Q (+ CIA, EAT-26; Appendix)}
};

\draw[arrow] (llm.south) -- (exp1.north);
\draw[arrow] (exp1) -- ($(dual.west)+(0,0.22)$);

\node[block=purple!6] (exp2) at (8, -3.2) {
    \textbf{Exp.\,II} Within-conv.\\[1pt]
    {\scriptsize 9 exchanges; assess at 3, 6, 9, c}
};
\draw[arrow] (llm.south) -- (exp2.north);
\draw[arrow] (exp2.east) -- ($(dual.west)+(0,-0.22)$);

\end{tikzpicture}%
}
\caption{Experimental design. Five ED vignettes $\times$ three prompt richness levels $\times$ six LLMs yield 90 conditions. Experiment~I tests between-conversation stability ($N\!=\!50$ independent runs per condition); Experiment~II tests within-conversation stability (9-exchange neutral dialogues, assessed at exchanges 3, 6, 9, with N = 20). Both use dual assessment: the persona completes self-report EDE-Q and three LLM raters from different providers complete the same instruments from the perspective of the individual based solely on the conversation transcript; secondary scales (CIA, EAT-26) are reported in Appendix~\ref{app:secondary}. Published assessment scores are excluded from all prompts.}
\label{fig:design}
\end{figure}
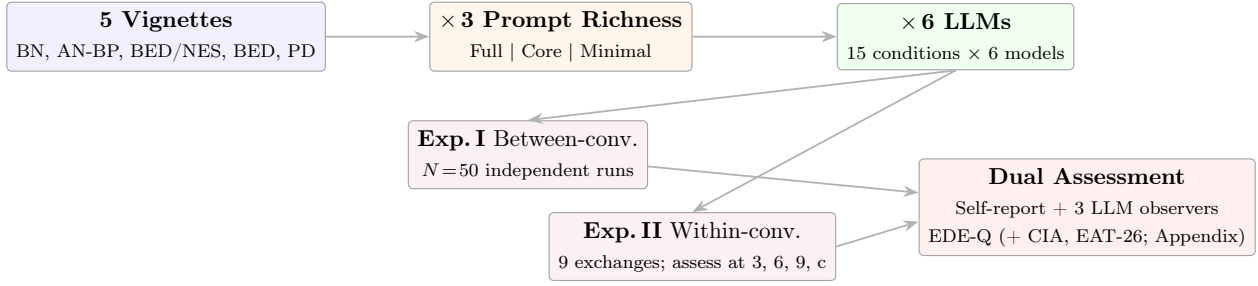

\subsection{Experimental design}

We operationalize stability through two complementary experiments (see Figure~\ref{fig:design}). \textit{Experiment~I} tests between-conversation stability: each model generates a first-person narrative (e.g., describing a typical day or eating-related situations) and completes ED self-report instruments; independent LLM raters complete the same self-report instruments from the perspective of the individual, based solely on the narrative and without access to persona instructions. \textit{Experiment~II} tests within-conversation stability: the target LLM engages in a 9-exchange neutral conversation (each exchange comprising one interviewer question and one persona response) with a scripted conversation partner who listens without judgment and asks simple questions that help the speaker continue talking without sharing opinions or trying to influence what they say. The conversation covers three topic blocks (the same three situations as Exp.~I, uniformly shuffled per run), with self-report and observer assessments triggered after each block (at exchanges 3, 6, and~9). Observer raters evaluate each block independently; $N = 20$ runs per condition.

\begin{table}[ht]
\centering
\caption{Large language models used in this study. Three models serve as both persona generators and observer-raters; three serve as generators only.}
\label{tab:LLMs}
\footnotesize
\begin{tabular}{llll}
\toprule
\textbf{Model} & \textbf{Provider} & \textbf{Role} & \textbf{Exp.} \\
\midrule
Claude Sonnet 4.6 & Anthropic & Gen + Obs & I + II \\
GPT 5.4           & OpenAI   & Gen + Obs & I + II \\
Gemini 3.1 Pro    & Google & Gen + Obs & I + II \\
DeepSeek V4 Flash & DeepSeek & Generator   & I + II \\
GPT OSS 120B      & Self-hosted & Generator   & I + II \\
Llama 3.3 70B     & Self-hosted & Generator   & I \\
\bottomrule
\end{tabular}
\end{table}

Six LLMs serve as persona generators (Table~\ref{tab:LLMs}); three also serve as observer-raters. Table~\ref{tab:info-access} (Appendix) details the information available to each role. Scores are averaged across the three raters. In Experiment~I, personas describe three eating-related situations in a single first-person narrative; in Experiment~II, the persona engages in a 9-exchange neutral dialogue (Appendix~\ref{app:task_prompts}). Variables: model (six in Exp.~I; five in Exp.~II; Llama excluded due to insufficient multi-turn capacity), persona case (five vignettes), and prompt richness (Full, Core, Minimal). Exp.~I uses 50 runs per condition (GPT~OSS: 35--49; Llama: 31--50 due to API instability); Exp.~II uses 20 runs with three within-conversation assessment timepoints.

\subsection{Persona construction and prompt richness}
\label{sec:ablation}

Each persona is conditioned via the system prompt, with an identical instruction prefix (``You are a patient with an eating disorder. Respond in first person, in character, at all times.'') followed by condition-specific clinical content. All API calls use default parameters with no top-p override. All published assessment scores are excluded from every prompt to prevent circularity. Two secondary scales (CIA, EAT-26) are co-administered for convergent validity (Appendix~\ref{app:secondary}).

We systematically vary prompt richness across three levels, operationalising the distinction in Fairburn's model between active maintaining mechanisms and predisposing/contextual factors:

\paragraph{Full vignette.}
The complete clinical case description as published, including demographics, illness history, onset narrative, family background, trauma history, social context, cognitive patterns, emotional states, behavioural symptoms, and substance use. The only omission is assessment scores.

\paragraph{Core maintaining mechanisms.}
Only the Fairburn transdiagnostic maintaining factors: over-evaluation of eating, shape, and weight (and their control); dietary restraint pattern and food rules; binge eating pattern and triggers; purging and compensatory behaviours; body checking and avoidance behaviours; and mood states that drive the binge--purge--restrict cycle. All biographical context, developmental history, family background, trauma, social relationships, and precipitating events are removed.

\paragraph{Minimal.}
Diagnostic label, age, gender, BMI, and quantified behavioural frequencies only (e.g., binge frequency, purge frequency, exercise pattern). No qualitative description of any kind.

This design tests three questions: (1)~does biographical and contextual richness improve persona accuracy and stability beyond what clinical information alone provides (Full vs.\ Core)? (2)~does qualitative clinical description add value beyond bare diagnostic parameters (Core vs.\ Minimal)? (3)~can LLMs produce clinically plausible ED behaviour from a diagnostic label and frequencies alone, potentially relying on stereotyped training-data representations (Minimal vs.\ ground truth)? Full prompt texts for all cases and conditions are provided in the supplementary material.

\subsection{Measurement and analysis}

The primary outcome is the \textbf{EDE-Q} (Eating Disorder Examination Questionnaire; 28 items, 0--6 scale; \citealt{fairburnEatingDisorderExamination2008}), with ground-truth scores available for all five cases. Two secondary measures, the CIA (Clinical Impairment Assessment; \citealt{bohnMeasurementImpairmentEating2008}) and EAT-26 (Eating Attitudes Test; \citealt{garnerEatingAttitudesTest1982}), are co-administered in every run for all personas; ground-truth scores are available for a subset of cases (CIA: BN, PD; EAT-26: AN-BP) and reported in Appendix~\ref{app:secondary}. All three instruments are administered simultaneously in a single prompt; responses are returned as structured JSON (one entry per item with questionnaire ID, item number, and numeric score) for automated parsing (Appendix~\ref{app:assessment_prompts}).

\paragraph{EDE-Q scoring.}
The EDE-Q comprises 28 items (0--6 scale); items~13--18 record raw behavioural counts and are excluded from scoring, in line with human-scoring practices. Four subscales are unweighted item means: Restraint (5 items), Eating Concern (5), Shape Concern (8), and Weight Concern (5); the global score is their mean. All subscale items are clamped to $[0, 6]$; four responses (item~19, Gemini) required clamping due to raw frequency counts.

For \textbf{self-report}, after the persona produces its narrative, the same model (same conversation context) completes all questionnaires based on subjective experience without access to ground-truth scores. For \textbf{observer assessment}, all three observer models (Claude, GPT~5.4, Gemini) independently rate every persona transcript ``from the perspective of the individual'' without access to the persona prompt or ground-truth scores. Scores are averaged across the three raters. Inter-rater agreement is poor to moderate: ICC(A,1)\,=\,.14--.43 across conditions (Exp.~I), consistent with systematic differences between observer models rather than purely random disagreement.

All analyses use Python~3 (\texttt{statsmodels}, \texttt{pingouin}, \texttt{scipy}). Variance decomposition uses Type~II ANOVA with model, case, and prompt richness as fixed factors; variance components are reported as percentages of total sum of squares. Within-conversation drift (Exp.~II) is tested per model via paired $t$-tests (exchange~3 vs.~9, paired by run). Each run is an independent API call with no shared state, justifying the treatment of runs as independent observations; analyses are stratified by model to avoid pooling across the nested design. Between-case discriminability is assessed via one-way ANOVA ($\eta^2$) per subscale. Where multiple tests address the same hypothesis family, $p$-values are Holm--Bonferroni adjusted. Inter-rater reliability uses ICC(A,1) (two-way random, absolute agreement; \citealt{shroutIntraclassCorrelationsUses1979}). Sample sizes ($N = 50$ Exp.~I; $N = 20$ Exp.~II) yield SE\,$\approx$\,0.03--0.04 at observed SD\,$\approx$\,0.2. Proprietary models were accessed via official APIs; open-weight models (GPT~OSS~120B, Llama~3.3~70B) were served with Ollama on NVIDIA H100 NVL GPU. Total inference time was 396 hours (207\,h Exp.~I, 189\,h Exp.~II) across approximately 37,900 API calls ($\sim$380M tokens).

\section{Results}
\label{sec:results}

Table~\ref{tab:scores} summarises EDE-Q global scores per case and model against published ground truth. We organise results around four questions: (1)~how stable are persona scores? (2)~does prompt richness matter? (3)~how accurate are scores relative to ground truth? (4)~what drives the accuracy mismatch?

\begin{table}[t]
\centering
\caption{EDE-Q global scores (0--6 scale): ground truth vs.\ LLM self-report per case and model (Full prompt).
All six models systematically overshoot ground truth; AN-BP scores above BN in four of six models despite lower ground-truth severity.}
\label{tab:scores}
\begin{threeparttable}
\begin{center}
\resizebox{0.7\textwidth}{!}{%
\begin{tabular}{lr rrrrrr}
\toprule
& & \multicolumn{6}{c}{\textbf{Self-report mean (Full prompt)}} \\
\cmidrule(lr){3-8}
\textbf{Case} & \textbf{GT} & \textbf{Claude} & \textbf{DeepSeek} & \textbf{GPT} & \textbf{GPT OSS} & \textbf{Gemini} & \textbf{Llama} \\
\midrule
BN       & 5.25 & 5.04 & 5.37 & 5.30 & 4.91 & 5.68 & 5.52 \\
AN-BP    & 4.30 & 5.40 & 5.62 & 5.50 & 4.86 & 5.97 & 5.51 \\
BED/NES  & 3.60 & 5.19 & 5.76 & 5.54 & 4.52 & 5.63 & 5.45 \\
BED      & 3.49 & 4.72 & 5.39 & 5.32 & 3.98 & 4.95 & 5.33 \\
PD       & 2.64 & 4.08 & 5.23 & 4.91 & 4.84 & 5.12 & 5.37 \\
\midrule
\multicolumn{2}{l}{\textit{Mean bias}} & +1.12 & +1.65 & +1.50 & +0.70 & +1.81 & +1.58 \\
\multicolumn{2}{l}{\textit{MAE}} & 1.18 & 1.65 & 1.51 & 0.91 & 1.81 & 1.59 \\
\multicolumn{2}{l}{\textit{CV (\%)}} & 3.2 & 3.6 & 2.3 & 7.7 & 3.1 & 3.3 \\
\multicolumn{2}{l}{\textit{Rank $\rho$}} & .60 & .60 & .60 & .00 & .60 & .60 \\
\bottomrule
\end{tabular}}%
\end{center}\smallskip
\begin{center}
\begin{minipage}{0.7\textwidth}%
\begin{tablenotes}[flushleft]
\small
\item Notes. GT = published ground-truth EDE-Q (0--6 scale). Bias = mean LLM $-$ GT across five cases. MAE = mean absolute error. CV = coefficient of variation (SD/mean, averaged across conditions). Rank~$\rho$ = Spearman correlation between GT and LLM severity ordering.
\end{tablenotes}
\end{minipage}
\end{center}
\end{threeparttable}
\end{table}

\subsection{Stability}
\label{sec:results_stability}

\begin{figure}[t]
\centering
\includegraphics[width=\textwidth]{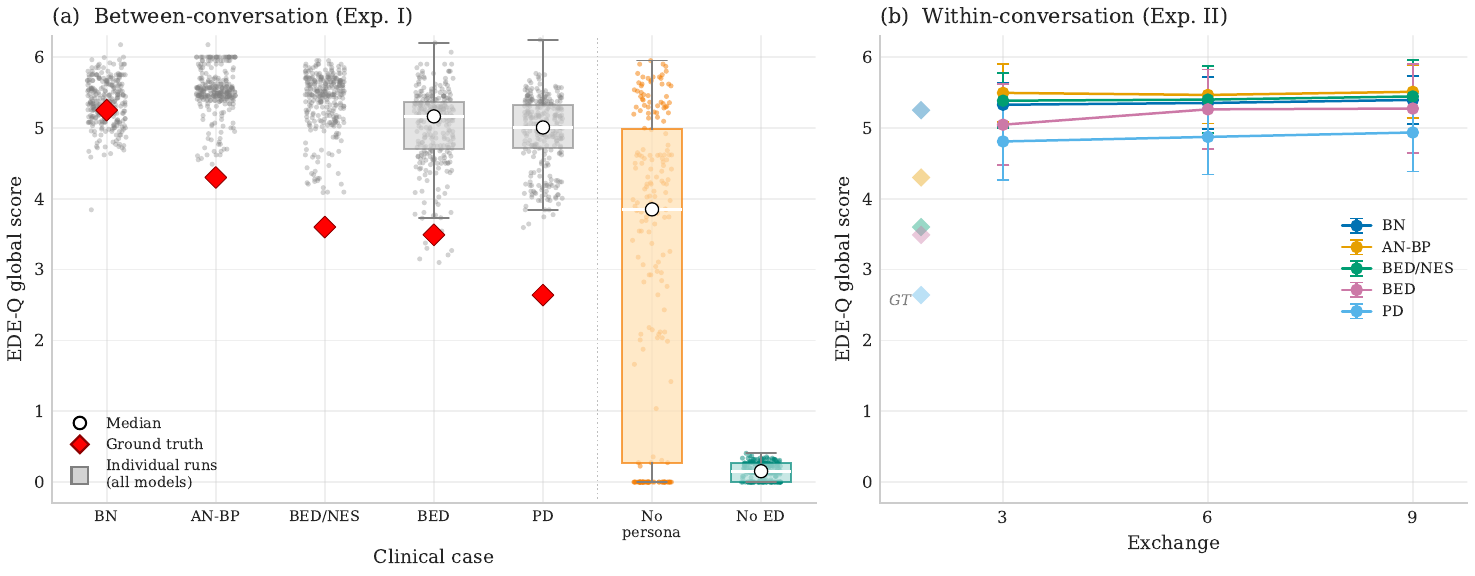}
\caption{Persona stability across and within conversations. \textbf{(a)}~Between-conversation (Exp.~I, Full prompt): each dot is one run's EDE-Q global score (all six models pooled); box plots show the median and IQR. Red diamonds mark published ground truth. Scores are tightly clustered (stable) but systematically above ground truth, with the gap increasing for lower-severity cases. Two controls (right of separator; see Appendix~\ref{app:controls} for detailed analysis): ``No persona'' shows model defaults without any persona prompt, spanning nearly the full scale due to between-model divergence (Appendix Table~\ref{tab:baseline}); ``No ED'' shows a healthy-persona control (four models, $N = 50$ each) with scores clustering near floor, confirming the overshoot is ED-specific (Appendix Table~\ref{tab:no-ed}). \textbf{(b)}~Within-conversation (Exp.~II, Full prompt): mean EDE-Q global score at exchanges 3, 6, and~9 ($\pm$1\,SD), per case (all five models pooled). Faded markers at left show ground truth. All cases drift upward; drift is concentrated in Restraint items (Appendix Table~\ref{tab:drift}).}
\label{fig:stability}
\end{figure}

\paragraph{Between-conversation stability (Exp.~I).}
Between-conversation stability is remarkably high (Figure~\ref{fig:stability}a). Coefficients of variation average 2--4\% for five of six models (GPT~OSS is an outlier at 7.7\%), and 30\% of all item--condition combinations produce the \emph{exact same score} across all 50 independent runs (SD\,=\,0). This near-deterministic responding under standard stochastic sampling is most pronounced for Weight Concern and Shape Concern items (42\% and 37\% zero-variance, respectively) and least for Eating Concern (18\%) and Restraint (23\%).

\paragraph{Within-conversation stability (Exp.~II).}
Experiment~II tests whether extending the conversation across a 9-exchange neutral dialogue introduces drift. Results are reported for five models (Claude, DeepSeek, GPT~5.4, GPT~OSS, and Gemini~3.1~Pro; $N = 20$ runs per condition).

First-assessment scores (exchange~3) replicate Experiment~I closely: across 20 model$\times$case cells (four models with matched Exp.~I data), Exp.~I and Exp.~II correlate at $r = .98$ ($p < .0001$), with a mean absolute difference of 0.08 EDE-Q points. The between-conversation patterns are therefore robust to procedural differences (single prompt vs.\ multi-turn dialogue). However, EDE-Q scores \emph{increase} over the conversation for four of five models (Figure~\ref{fig:stability}b; Appendix Table~\ref{tab:drift}): global drift ranges from $+0.08$ (DeepSeek) to $+0.22$ (Gemini); all $p_\text{adj} \leq .020$ after Holm--Bonferroni correction. GPT~OSS is the exception, drifting \emph{downward} ($\Delta = -0.09$, n.s.). The upward drift is statistically significant but clinically small, and concentrated in Restraint ($+0.21$ to $+0.74$) rather than Shape Concern ($+0.01$ to $+0.11$) or Weight Concern (n.s.), consistent with the ceiling pattern from Exp.~I.

\paragraph{Self--observer divergence over exchanges.}
Self-report scores drift upward for four of five models; GPT~OSS is the exception, drifting slightly downward. Observer scores show model-specific patterns: GPT~5.4 observers rate successive blocks \emph{lower} (widening self--observer gap), Gemini and DeepSeek observers increase in parallel with self-report, and GPT~OSS observers decline faster than self-report, reversing an initial $-0.20$ gap (observers higher) to $+0.12$ (self-report higher) by exchange~9.

\subsection{Effect of persona description richness}
\label{sec:results_ablation}

A Type~II ANOVA decomposes EDE-Q global score variance into four sources (Appendix Table~\ref{tab:variance}): model (49\%), case (14\%), prompt richness (1\%), and residual (35\%). Prompt richness contributes negligibly, suggesting that even stripped-down persona descriptions (Minimal: diagnostic label, age, gender, BMI, and behavioural frequencies only) elicit near-equivalent responding to full clinical vignettes. Model identity is the strongest single predictor, consistent with the large between-model differences in absolute severity (Gemini\,$>$\,DeepSeek\,$>$\,Llama\,$>$\,GPT\,$>$\,Claude\,$>$\,GPT~OSS).

\subsection{Accuracy relative to ground truth}
\label{sec:results_accuracy}

\begin{figure}[t]
\centering
\includegraphics[width=\textwidth]{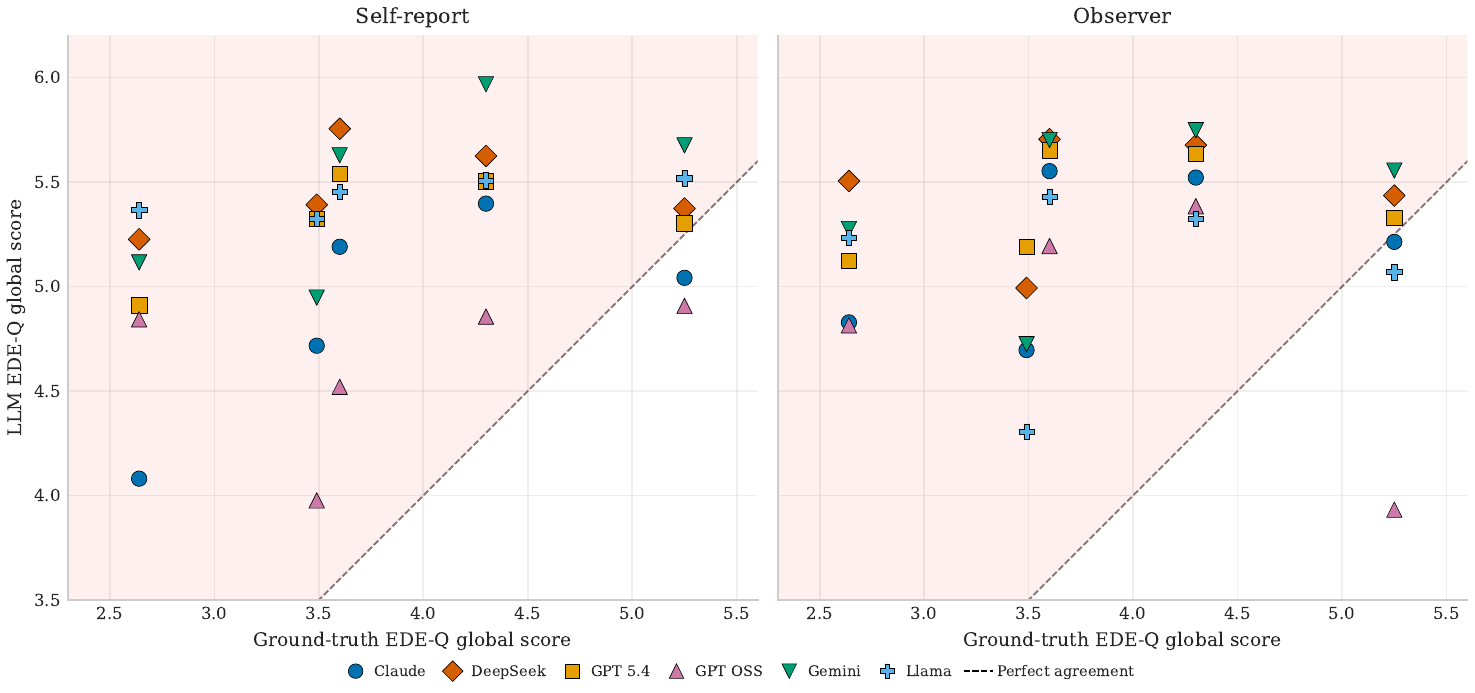}
\caption{LLM mean EDE-Q global score versus ground truth (Full prompt), shown separately for self-report (left) and observer ratings (right). The dashed line marks perfect agreement; the shaded region highlights the overshoot zone. All points lie above the identity line, compressed into a narrow band despite a ground-truth range of 2.64--5.25. AN-BP exceeds BN in four of six models despite lower ground-truth severity.}
\label{fig:accuracy}
\end{figure}

\paragraph{Systematic overshoot with severity-dependent compression.}
All six models overshoot ground-truth EDE-Q scores, with mean bias ranging from $+0.70$ (GPT~OSS) to $+1.81$ (Gemini; Table~\ref{tab:scores}). The overshoot is severity-dependent: BN (GT\,=\,5.25) is reproduced within $-0.34$ to $+0.43$ points, while PD (GT\,=\,2.64) is inflated by $+1.44$ to $+2.73$ points. Critically, the LLM rank ordering of cases does not match ground truth: AN-BP scores \emph{higher} than BN in four of six models, and BED/NES exceeds BN in three, despite both having lower ground-truth severity (Figure~\ref{fig:accuracy}). Only GPT~OSS and Llama correctly rank BN above AN-BP and BED/NES, but GPT~OSS achieves zero overall rank-order accuracy (Spearman $\rho = 0$). Additional conversational context does not help: Exp.~II shows that scores drift \emph{further} from ground truth over the dialogue for four of five models (Section~\ref{sec:results_stability}).

\paragraph{Self-report versus observer agreement.}
Self-report and observer ratings show the same overshoot pattern: mean self-report EDE-Q is 5.20 versus 5.15 for observers (Appendix Table~\ref{tab:self-vs-obs}; Appendix Figure~\ref{fig:self_vs_obs}). Both sources show the same rank inversions and the same range compression, indicating that the generated conversation already encodes the exaggerated severity and observers faithfully rate what they see (see Section~\ref{sec:discussion}). The direction of the self--observer gap varies by model: Claude's observers rate \emph{higher} than self-report ($\Delta = -0.28$), while Llama and GPT~OSS show the opposite ($\Delta = +0.36$ and $+0.23$, respectively). A systematic judge-level bias is also evident: Gemini-as-judge rates 0.43 points higher than GPT~5.4-as-judge on average across all generators (Appendix Table~\ref{tab:self-vs-obs}).

\paragraph{Secondary scales.}
Two co-administered instruments confirm the overshoot is not specific to the EDE-Q (Appendix Table~\ref{tab:secondary}). The CIA (Clinical Impairment Assessment; 0--48 scale), which measures functional consequences of eating pathology, shows bias of $+0.9$ to $+10.5$ for the two cases with ground truth (BN and PD; both GT\,=\,31). The EAT-26 (Eating Attitudes Test; 0--78 scale; GT for AN-BP\,=\,35) shows high between-model variability, with bias ranging from $-13.7$ (Llama) to $+19.9$ (Gemini).

\subsection{Selective stereotyping: why models overshoot}
\label{sec:results_mismatch}

The overshoot and rank inversions are not random errors; they follow a systematic pattern that we term \emph{selective stereotyping}. Decomposing the EDE-Q into its four subscales (Table~\ref{tab:subscales}; Appendix Figure~\ref{fig:subscales}) reveals that models treat eating disorders as a two-component construct and handle each component differently:

\begin{table}[tbp]
\centering
\small
\caption{EDE-Q subscale means across all six models (Full prompt). Cognitive-affective subscales (SC, WC) are near-saturated regardless of case severity; only the behavioural subscale (Restraint) differentiates cases. This selective stereotyping (maximal body dissatisfaction for any ED diagnosis) drives both the systematic overshoot and the rank inversions.}
\label{tab:subscales}
\begin{tabular}{lrrrrr}
\toprule
& \textbf{Restraint} & \textbf{EC} & \textbf{SC} & \textbf{WC} & \textbf{Global} \\
\midrule
BN (GT\,=\,5.25)       & 5.27 & 4.96 & 5.48 & 5.52 & 5.31 \\
AN-BP (GT\,=\,4.30)    & 5.42 & 5.21 & 5.65 & 5.65 & 5.48 \\
BED/NES (GT\,=\,3.60)  & 5.10 & 5.08 & 5.64 & 5.64 & 5.36 \\
BED (GT\,=\,3.49)      & 3.44 & 5.30 & 5.61 & 5.57 & 4.98 \\
PD (GT\,=\,2.64)    & 4.67 & 4.53 & 5.32 & 5.17 & 4.92 \\
\midrule
\textit{Between-case range}     & \textit{1.98} & \textit{0.77} & \textit{0.33} & \textit{0.48} & \textit{0.56} \\
\bottomrule
\end{tabular}
\end{table}

\paragraph{Behavioural items respond to case content.}
Restraint---the only subscale measuring concrete dietary behaviours---differentiates cases with a between-case range of 1.98 points. BED shows low Restraint (3.44), consistent with binge eating without restriction; AN-BP shows the highest (5.42), consistent with extreme dietary rules. The variance decomposition confirms this (Appendix Table~\ref{tab:variance}): case explains 47\% of Restraint variance.

\paragraph{Cognitive-affective items default to ceiling.}
Shape Concern and Weight Concern are near-saturated regardless of case severity (between-case ranges of only 0.33 and 0.48). Case explains just 3\% of SC variance; model identity explains 66\%. Models appear to encode a default: \emph{any eating disorder entails maximal body dissatisfaction}. This is clinically incorrect: moderate presentations (e.g., PD, GT\,=\,2.64) involve substantially less weight/shape preoccupation.

\paragraph{Ceiling saturation explains the rank inversions.}
With SC and WC at ceiling for all cases, the global score is driven almost entirely by Restraint. AN-BP (Restraint\,=\,5.42) therefore exceeds BN (5.27) despite lower ground-truth severity, because its vignette describes extreme dietary restriction while BN emphasises purging. The rank ordering reflects \emph{diagnosis type} (how much the patient restricts) rather than \emph{overall severity}.

\section{Discussion}
\label{sec:discussion}

We asked whether LLMs can maintain coherent, clinically accurate eating-disorder personas across repeated assessments. The answer is qualified: personas are remarkably stable (CV\,=\,2--4\%) yet systematically stereotyped. Behavioural Restraint tracks case content, but cognitive-affective items---Shape Concern, Weight Concern---default to ceiling regardless of severity, encoding a lay representation (``any eating disorder entails maximal body dissatisfaction'') rather than the clinically graded construct. This is not mere instruction-following: observer models that never see the persona prompt replicate the same overshoot and rank inversions (Appendix Table~\ref{tab:self-vs-obs}); prompt richness explains only 1\% of variance (Appendix Table~\ref{tab:variance}); and 30\% of item responses are deterministic at ceiling across all runs. The stability we document is stability of \emph{bias}, a generative prior constrained but not determined by the prompt. Our findings replicate and extend \citet{gonnermann-mullerStablePersonasDualAssessment2026}, who report stable ADHD self-reports under dual assessment; we show that validated ground-truth scores reveal what stability alone cannot, namely that consistent self-reports may reflect consistent \emph{inaccuracy} rather than consistent fidelity. The pattern also converges with \citet{khadangiWhenAITakes2025}, whose therapy-style prompting pushes frontier models into ``multi-morbid synthetic psychopathology''; here the mechanism is the same (training-data priors compressing the clinical spectrum into its extreme end) but manifested through structured self-report rather than open dialogue. The stereotypy varies across models (inter-case profile correlations from $r = .39$ for Gemini to $r = .85$ for Llama) yet is unaffected by prompt richness, confirming its origin in pretraining rather than insufficient context. Preliminary cross-instrument evidence supports this interpretation: the Clinical Impairment Assessment (CIA), which measures functional consequences of eating pathology rather than body-image cognitions, shows substantially less overshoot for the moderate case (PD: 8\% of scale range vs.\ 38\% on the EDE-Q; Appendix Table~\ref{tab:secondary}). When the instrument does not directly probe body dissatisfaction, the ceiling effect largely disappears. Even the healthy-persona control retains faint traces of this asymmetry: Shape Concern carries residual signal (up to 0.84) while Restraint and Eating Concern are exactly zero (Appendix Table~\ref{tab:no-ed}). Whether this reflects embedded training-data associations or a normatively plausible degree of body dissatisfaction for a young woman is ambiguous, but the cognitive-affective $>$ behavioural ordering mirrors the ED pattern in miniature.

Two technical features rule out simple remediation. First, the non-monotonic rank inversions (Section~\ref{sec:results_accuracy})---AN-BP scoring above BN despite lower ground-truth severity---mean no monotonic transform can recalibrate the outputs; the distortion originates at the subscale level (Section~\ref{sec:results_mismatch}), where ceiling-saturated SC/WC items leave the global score driven by Restraint, which tracks \emph{diagnosis type} rather than \emph{overall severity}. Second, within-conversation scores drift \emph{upward} for four of five models (Exp.~II), moving further from ground truth rather than converging on it; more dialogue does not help. A caveat is warranted: ground-truth scores are single expert administrations with test--retest reliability of $r \approx .81$--.94 \citep{luceReliabilityEatingDisorder1999}, not objective benchmarks. Yet the overshoot magnitude (0.7--1.8 points on a 0--6 scale) and its systematic direction (always above GT, never below) far exceed what measurement noise alone can explain. Two control conditions bracket the interpretation. Without any persona, model defaults span 0.17--5.45 on the global scale (Appendix Table~\ref{tab:baseline}), but this baseline is confounded by the food-focused task prompt, which primes models to generate clinically relevant content even without clinical instruction. A cleaner comparison is the healthy-persona control (``No ED''; Appendix Table~\ref{tab:no-ed}, Figure~\ref{fig:stability}a), which yields near-floor scores across all four models (global 0.00--0.27 self-report, 0.02--0.08 observer), confirming that models \emph{can} represent low pathology when instructed on the same task. The overshoot is therefore not a task artifact, not a general tendency to endorse symptoms, and not a failure to produce low scores; it is specific to the models' representation of graded eating-disorder severity (see Appendix~\ref{app:controls} for detailed analysis).

The practical consequence is a severity-specific failure we term the ``missing middle.'' Severe cases such as BN (GT\,=\,5.25) are reproduced within $\pm 0.43$ points, but moderate presentations (GT\,=\,2.6--3.6) are inflated by 0.5--2.7 points, precisely the cases most relevant for clinical training. Aggregate validation can therefore appear adequate while masking failures where fidelity matters most. Because the LLM-to-GT mapping is non-monotonic, post-hoc calibration is not viable; what is needed is item-level clinical grounding at generation time and representational diversity that current training data does not provide.

\section{Limitations}
\label{sec:limitations}

Exp.~II uses fewer runs ($N = 20$ vs.\ 50) and five of six models (Llama excluded due to insufficient multi-turn capacity); the 9-exchange dialogue may be too short to capture drift in extended interactions. Personas derive from published case vignettes, not real patient interactions, though a licensed clinical psychologist on the author team confirmed that generated narratives are qualitatively realistic despite the quantitative overshoot. All five vignettes feature female patients, reflecting ED case study publication bias. We lack human baselines (clinicians or actors role-playing the same personas), so whether the overshoot is model-specific or inherent to self-report simulation remains open. Ground-truth scores are single-administration point estimates; human test--retest variability ($r \approx .81$--.94; \citealt{luceReliabilityEatingDisorder1999}) means some LLM--GT discrepancy may fall within measurement error. The no-persona and healthy-persona controls (Appendix~\ref{app:controls}) use four of six models; whether GPT~OSS and Llama show the same patterns remains untested. Whether selective stereotyping generalises beyond eating disorders remains untested; disorders with behaviourally graded severity at all levels (e.g., PHQ-9) may show less ceiling saturation.

\section{Conclusion}
\label{sec:conclusion}

Across six LLMs and five ED personas, we find a paradox: LLM personas are \emph{too stable and too inaccurate}. Between-conversation variability is negligible (CV\,$\approx$\,2--4\%), yet scores systematically overshoot ground truth by 12--30\%. The mechanism is selective stereotyping: cognitive-affective items saturate at ceiling regardless of case severity while only behavioural restraint differentiates cases, producing rank inversions no calibration can correct. Dual assessment confirms this is a generation problem: self-report and observer ratings show identical patterns. LLMs can portray severe eating pathology but lack a representation of moderate pathology, the ``missing middle.'' Improving fidelity requires item-level clinical grounding, not richer prompts or post-hoc calibration. Key next steps are cross-disorder replication with behaviourally graded instruments (e.g., PHQ-9) and human baseline studies to establish whether the overshoot is model-specific or inherent to simulated self-report.

\section*{Acknowledgments}
Research reported in this paper was partially supported through the Research Campus Modal funded by the German Federal Ministry of Research, Technology and Space (BMFTR) (fund numbers 05M14ZAM,05M20ZBM) and the German Research Foundation (DFG) through the DFG Cluster of Excellence MATH+ (EXC-2046/1, project ID 390685689, project AA3-15), by the German Federal Ministry of Research, Technology and Space (BMFTR), grant number 16DII133 (Weizenbaum-Institute), by the German Research Foundation (DFG), CRC 1404: “FONDA: Foundations of Workflows for Large-Scale Scientific Data Analysis” (Project-ID 414984028) and by the Zuse Institute Berlin via RISE@ZIB services for hosting the LLM models.

\bibliographystyle{plainnat}
\bibliography{bibliography}

\begin{thebibliography}{45}
\providecommand{\natexlab}[1]{#1}
\providecommand{\url}[1]{\texttt{#1}}
\expandafter\ifx\csname urlstyle\endcsname\relax
  \providecommand{\doi}[1]{doi: #1}\else
  \providecommand{\doi}{doi: \begingroup \urlstyle{rm}\Url}\fi

\bibitem[Abdulhai et~al.(2025)Abdulhai, Cheng, Clay, Althoff, Levine, and Jaques]{abdulhaiConsistentlySimulatingHuman2025}
Marwa Abdulhai, Ryan Cheng, Donovan Clay, Tim Althoff, Sergey Levine, and Natasha Jaques.
\newblock Consistently {{Simulating Human Personas}} with {{Multi-Turn Reinforcement Learning}}, October 2025.

\bibitem[Acevedo et~al.(2026)Acevedo, Aneja, Opler, Valera, and Jarmon]{acevedoEvaluatingEfficacyChatGPT352026}
Sebastian Acevedo, Esha Aneja, Douglas~J. Opler, Pamela Valera, and Eric Jarmon.
\newblock Evaluating the {{Efficacy}} of {{ChatGPT-3}}.5 {{Versus Human-Delivered Text-Based Cognitive-Behavioral Therapy}}: {{A Comparative Pilot Study}}.
\newblock \emph{American Journal of Psychotherapy}, 79\penalty0 (1):\penalty0 4--11, March 2026.
\newblock ISSN 0002-9564, 2575-6559.
\newblock \doi{10.1176/appi.psychotherapy.20240070}.

\bibitem[Ajluni(2025)]{ajluniArtificialIntelligencePsychiatric2025}
Victor Ajluni.
\newblock Artificial intelligence in psychiatric education: {{Enhancing}} clinical competence through simulation.
\newblock \emph{Industrial Psychiatry Journal}, 34\penalty0 (1):\penalty0 11, January 2025.
\newblock ISSN 0972-6748.
\newblock \doi{10.4103/ipj.ipj_377_24}.

\bibitem[Beck et~al.(2024)Beck, Schuff, Lauscher, and Gurevych]{beckSensitivityPerformanceRobustness2024}
Tilman Beck, Hendrik Schuff, Anne Lauscher, and Iryna Gurevych.
\newblock Sensitivity, {{Performance}}, {{Robustness}}: {{Deconstructing}} the {{Effect}} of {{Sociodemographic Prompting}}.
\newblock In Yvette Graham and Matthew Purver, editors, \emph{Proceedings of the 18th {{Conference}} of the {{European Chapter}} of the {{Association}} for {{Computational Linguistics}} ({{Volume}} 1: {{Long Papers}})}, pages 2589--2615, St. Julian's, Malta, March 2024. Association for Computational Linguistics.
\newblock \doi{10.18653/v1/2024.eacl-long.159}.

\bibitem[Bohn et~al.(2008)Bohn, Doll, Cooper, O'Connor, Palmer, and Fairburn]{bohnMeasurementImpairmentEating2008}
Kristin Bohn, Helen~A. Doll, Zafra Cooper, Marianne O'Connor, Robert~L. Palmer, and Christopher~G. Fairburn.
\newblock The measurement of impairment due to eating disorder psychopathology.
\newblock \emph{Behaviour Research and Therapy}, 46\penalty0 (10):\penalty0 1105--1110, 2008.

\bibitem[Cummins(2026)]{cumminsThreatAnalyticFlexibility2026}
Jamie Cummins.
\newblock The threat of analytic flexibility in using large language models to simulate human data, April 2026.

\bibitem[Denecke et~al.(2026)Denecke, {Rivera-Romero}, {L{\'o}pez-Campos}, Dorronzoro, and Gabarron]{deneckeUncoveringAIsHidden2026}
Kerstin Denecke, Octavio {Rivera-Romero}, Guillermo {L{\'o}pez-Campos}, Enrique Dorronzoro, and Elia Gabarron.
\newblock Uncovering {{AI}}'s hidden risks: An empirical analysis of health-related {{AI}} incidents and their ethical implications.
\newblock \emph{AI and Ethics}, 6\penalty0 (2):\penalty0 169, February 2026.
\newblock ISSN 2730-5961.
\newblock \doi{10.1007/s43681-026-01012-7}.

\bibitem[{Erg{\"u}ney-Okumu{\c s}}(2023)]{erguney-okumusIntegratingEMDREnhanced2023}
F.~Elif {Erg{\"u}ney-Okumu{\c s}}.
\newblock Integrating {{EMDR With Enhanced Cognitive Behavioral Therapy}} in the {{Treatment}} of {{Bulimia Nervosa}}: {{A Single Case Study}}.
\newblock \emph{Journal of EMDR Practice and Research}, 15\penalty0 (4):\penalty0 231--243, May 2023.
\newblock \doi{10.1891/EMDR-D-21-00012}.

\bibitem[Fairburn and Beglin(2008)]{fairburnEatingDisorderExamination2008}
Christopher~G. Fairburn and Sarah~J. Beglin.
\newblock Eating disorder examination questionnaire.
\newblock In \emph{Cognitive Behavior Therapy and Eating Disorders}, pages 309--313. Guilford Press, 2008.

\bibitem[Fairburn et~al.(2003)Fairburn, Cooper, and Shafran]{fairburnTransdiagnosticTheoryEating2003}
Christopher~G. Fairburn, Zafra Cooper, and Roz Shafran.
\newblock Cognitive behaviour therapy for eating disorders: A ``transdiagnostic'' theory and treatment.
\newblock \emph{Behaviour Research and Therapy}, 41\penalty0 (5):\penalty0 509--528, 2003.
\newblock \doi{10.1016/S0005-7967(02)00088-8}.

\bibitem[Frost et~al.(2025)Frost, Strodl, and Akosile]{frostMetaemotionTherapyComplex2025}
Gemma Frost, Esben Strodl, and Wole Akosile.
\newblock Meta-emotion therapy for complex trauma and binge eating: {{A}} case study.
\newblock \emph{Psychological Trauma: Theory, Research, Practice and Policy}, 17\penalty0 (4):\penalty0 904--911, May 2025.
\newblock ISSN 1942-969X.
\newblock \doi{10.1037/tra0001675}.

\bibitem[Garner et~al.(1982)Garner, Olmsted, Bohr, and Garfinkel]{garnerEatingAttitudesTest1982}
David~M. Garner, Marion~P. Olmsted, Yvonne Bohr, and Paul~E. Garfinkel.
\newblock The eating attitudes test: Psychometric features and clinical correlates.
\newblock \emph{Psychological Medicine}, 12\penalty0 (4):\penalty0 871--878, 1982.

\bibitem[{Gonnermann-M{\"u}ller} et~al.(2026){Gonnermann-M{\"u}ller}, Haase, Leins, Kosch, and Pokutta]{gonnermann-mullerStablePersonasDualAssessment2026}
Jana {Gonnermann-M{\"u}ller}, Jennifer Haase, Nicolas Leins, Thomas Kosch, and Sebastian Pokutta.
\newblock Stable {{Personas}}: {{Dual-Assessment}} of {{Temporal Stability}} in {{LLM-Based Human Simulation}}, January 2026.

\bibitem[Hambleton et~al.(2020)Hambleton, Hanstock, Simeone, and Sperling]{hambletonGroupDeliveredEnhancedCognitive2020}
Ashlea~L. Hambleton, Tanya~L. Hanstock, Rachel Simeone, and Michelle Sperling.
\newblock Group-{{Delivered Enhanced Cognitive Behavior Therapy}}: {{A Focus}} on a {{Young Adult Woman With Bulimia Nervosa}}.
\newblock \emph{Clinical Case Studies}, 19\penalty0 (1):\penalty0 62--77, February 2020.
\newblock ISSN 1534-6501.
\newblock \doi{10.1177/1534650119886653}.

\bibitem[Huang et~al.(2024)Huang, Zhang, Soto, and Evans]{huangDesigningLLMAgentsPersonalities2024}
Muhua Huang, Xijuan Zhang, Christopher Soto, and James Evans.
\newblock Designing {{LLM-Agents}} with {{Personalities}}: {{A Psychometric Approach}}, October 2024.

\bibitem[Khadangi et~al.(2025)Khadangi, Marxen, Sartipi, Tchappi, and Fridgen]{khadangiWhenAITakes2025}
Afshin Khadangi, Hanna Marxen, Amir Sartipi, Igor Tchappi, and Gilbert Fridgen.
\newblock When {{AI}} takes the couch: {{Psychometric}} jailbreaks reveal internal conflict in frontier models, December 2025.

\bibitem[Li et~al.(2024)Li, Huang, Wang, Zhang, Zou, and Sun]{liQuantifyingAIPsychology2024}
Yuan Li, Yue Huang, Hongyi Wang, Xiangliang Zhang, James Zou, and Lichao Sun.
\newblock Quantifying {{AI Psychology}}: {{A Psychometrics Benchmark}} for {{Large Language Models}}, June 2024.

\bibitem[Liu et~al.(2025)Liu, Jain, Takuri, Vege, Akalin, Zhu, O'Brien, and Sharma]{liuTRUTHDECAYQuantifying2025}
Joshua Liu, Aarav Jain, Soham Takuri, Srihan Vege, Aslihan Akalin, Kevin Zhu, Sean O'Brien, and Vasu Sharma.
\newblock {{TRUTH DECAY}}: {{Quantifying Multi-Turn Sycophancy}} in {{Language Models}}, February 2025.

\bibitem[Lozoya et~al.(2025)Lozoya, Conway, Duro, and D'Alfonso]{lozoyaLeveragingLargeLanguage2025}
Daniel~Cabrera Lozoya, Mike Conway, Edoardo Sebastiano~De Duro, and Simon D'Alfonso.
\newblock Leveraging {{Large Language Models}} for {{Simulated Psychotherapy Client Interactions}}: {{Development}} and {{Usability Study}} of {{Client101}}.
\newblock \emph{JMIR Medical Education}, 11\penalty0 (1):\penalty0 e68056, July 2025.
\newblock \doi{10.2196/68056}.

\bibitem[Luce and Crowther(1999)]{luceReliabilityEatingDisorder1999}
Kristine~H. Luce and Janis~H. Crowther.
\newblock The reliability of the {{Eating Disorder Examination}}---self-report questionnaire version ({{EDE-Q}}).
\newblock \emph{International Journal of Eating Disorders}, 25\penalty0 (3):\penalty0 349--351, 1999.
\newblock \doi{10.1002/(SICI)1098-108X(199904)25:3<349::AID-EAT15>3.0.CO;2-M}.

\bibitem[{Luz de Araujo} et~al.(2025){Luz de Araujo}, R{\"o}ttger, Hovy, and Roth]{luzdearaujoPrincipledPersonasDefining2025}
Pedro~Henrique {Luz de Araujo}, Paul R{\"o}ttger, Dirk Hovy, and Benjamin Roth.
\newblock Principled {{Personas}}: {{Defining}} and {{Measuring}} the {{Intended Effects}} of {{Persona Prompting}} on {{Task Performance}}.
\newblock In Christos Christodoulopoulos, Tanmoy Chakraborty, Carolyn Rose, and Violet Peng, editors, \emph{Proceedings of the 2025 {{Conference}} on {{Empirical Methods}} in {{Natural Language Processing}}}, pages 26857--26886, Suzhou, China, November 2025. Association for Computational Linguistics.
\newblock ISBN 979-8-89176-332-6.
\newblock \doi{10.18653/v1/2025.emnlp-main.1364}.

\bibitem[Manning et~al.(2024)Manning, Zhu, and Horton]{manningAutomatedSocialScience2024}
Benjamin~S. Manning, Kehang Zhu, and John~J. Horton.
\newblock Automated {{Social Science}}: {{Language Models}} as {{Scientist}} and {{Subjects}}, April 2024.

\bibitem[Masuda et~al.(2016)Masuda, Ng, Moore, Felix, and Drake]{masudaAcceptanceCommitmentTherapy2016}
Akihiko Masuda, Stacey~Y. Ng, Makeda Moore, Isabelle Felix, and Chad~E. Drake.
\newblock Acceptance and commitment therapy as a treatment for a {{Latina}} young adult woman with purging: {{A}} case report.
\newblock \emph{Practice Innovations}, 1\penalty0 (1):\penalty0 20--35, 2016.
\newblock ISSN 2377-8903.
\newblock \doi{10.1037/pri0000012}.

\bibitem[Melisse and Arora(2024)]{melisseCognitiveBehavioralTherapyenhanced2024}
Bernou Melisse and Teresa Arora.
\newblock Cognitive behavioral therapy-enhanced through videoconferencing for night eating syndrome, binge-eating disorder and comorbid insomnia: A {{Case Report}}.
\newblock \emph{Journal of Eating Disorders}, 12\penalty0 (1):\penalty0 175, November 2024.
\newblock ISSN 2050-2974.
\newblock \doi{10.1186/s40337-024-01131-8}.

\bibitem[Mitsopoulos et~al.(2023)Mitsopoulos, Bose, Mather, Bhatia, Gluck, Dorr, Lebiere, and Pirolli]{mitsopoulosPsychologicallyValidGenerativeAgents2023}
Konstantinos Mitsopoulos, Ritwik Bose, Brodie Mather, Archna Bhatia, Kevin Gluck, Bonnie Dorr, Christian Lebiere, and Peter Pirolli.
\newblock Psychologically-{{Valid Generative Agents}}: {{A Novel Approach}} to {{Agent-Based Modeling}} in {{Social Sciences}}.
\newblock \emph{Proceedings of the AAAI Symposium Series}, 2\penalty0 (1):\penalty0 340--348, 2023.
\newblock ISSN 2994-4317.
\newblock \doi{10.1609/aaaiss.v2i1.27698}.

\bibitem[Ozgun et~al.(2025)Ozgun, Pei, Hindriks, Donatelli, Liu, and Wang]{ozgunTrustworthyAIPsychotherapy2025}
Mithat~Can Ozgun, Jiahuan Pei, Koen Hindriks, Lucia Donatelli, Qingzhi Liu, and Junxiao Wang.
\newblock Trustworthy {{AI Psychotherapy}}: {{Multi-Agent LLM Workflow}} for {{Counseling}} and {{Explainable Mental Disorder Diagnosis}}, August 2025.

\bibitem[Paglieri et~al.(2026)Paglieri, Cross, Cunningham, Leibo, and Vezhnevets]{paglieriPersonaGeneratorsGenerating2026}
Davide Paglieri, Logan Cross, William~A. Cunningham, Joel~Z. Leibo, and Alexander~Sasha Vezhnevets.
\newblock Persona {{Generators}}: {{Generating Diverse Synthetic Personas}} at {{Scale}}, February 2026.

\bibitem[Panickssery et~al.(2024)Panickssery, Bowman, and Feng]{panicksseryLLMEvaluatorsRecognize2024}
Arjun Panickssery, Samuel~R. Bowman, and Shi Feng.
\newblock {{LLM Evaluators Recognize}} and {{Favor Their Own Generations}}.
\newblock \emph{Advances in Neural Information Processing Systems}, 37:\penalty0 68772--68802, December 2024.
\newblock \doi{10.52202/079017-2197}.

\bibitem[Park et~al.(2023)Park, O'Brien, Cai, Morris, Liang, and Bernstein]{parkGenerativeAgentsInteractive2023}
Joon~Sung Park, Joseph O'Brien, Carrie~Jun Cai, Meredith~Ringel Morris, Percy Liang, and Michael~S. Bernstein.
\newblock Generative {{Agents}}: {{Interactive Simulacra}} of {{Human Behavior}}.
\newblock In \emph{Proceedings of the 36th {{Annual ACM Symposium}} on {{User Interface Software}} and {{Technology}}}, {{UIST}} '23, pages 1--22, New York, NY, USA, October 2023. Association for Computing Machinery.
\newblock ISBN 979-8-4007-0132-0.
\newblock \doi{10.1145/3586183.3606763}.

\bibitem[Park et~al.(2024)Park, Zou, Shaw, Hill, Cai, Morris, Willer, Liang, and Bernstein]{parkGenerativeAgentSimulations2024}
Joon~Sung Park, Carolyn~Q. Zou, Aaron Shaw, Benjamin~Mako Hill, Carrie Cai, Meredith~Ringel Morris, Robb Willer, Percy Liang, and Michael~S. Bernstein.
\newblock Generative {{Agent Simulations}} of 1,000 {{People}}, November 2024.

\bibitem[Salewski et~al.(2023)Salewski, Alaniz, {Rio-Torto}, Schulz, and Akata]{salewskiInContextImpersonationReveals2023}
Leonard Salewski, Stephan Alaniz, Isabel {Rio-Torto}, Eric Schulz, and Zeynep Akata.
\newblock In-{{Context Impersonation Reveals Large Language Models}}' {{Strengths}} and {{Biases}}.
\newblock \emph{Advances in Neural Information Processing Systems}, 36:\penalty0 72044--72057, December 2023.

\bibitem[Samuel et~al.(2025)Samuel, Zou, Zhou, Chaudhari, Kalyan, Rajpurohit, Deshpande, Narasimhan, and Murahari]{samuelPersonagymEvaluatingPersona2024}
Vinay Samuel, Henry~Peng Zou, Yue Zhou, Shreyas Chaudhari, Ashwin Kalyan, Tanmay Rajpurohit, Ameet Deshpande, Karthik Narasimhan, and Vishvak Murahari.
\newblock Personagym: {{Evaluating}} persona agents and {{LLMs}}.
\newblock In \emph{Findings of the {{Association}} for {{Computational Linguistics}}: {{EMNLP}} 2025}, pages 6999--7022, Suzhou, China, November 2025. Association for Computational Linguistics.
\newblock \doi{10.18653/v1/2025.findings-emnlp.368}.

\bibitem[Sclar et~al.(2024)Sclar, Choi, Tsvetkov, and Suhr]{sclarQuantifyingLanguageModels2024}
Melanie Sclar, Yejin Choi, Yulia Tsvetkov, and Alane Suhr.
\newblock Quantifying {{Language Models}}' {{Sensitivity}} to {{Spurious Features}} in {{Prompt Design}} or: {{How I}} learned to start worrying about prompt formatting.
\newblock In \emph{The Twelfth International Conference on Learning Representations}, 2024.
\newblock URL \url{https://openreview.net/forum?id=RIu5lyNXjT}.

\bibitem[Sharma et~al.(2024)Sharma, Tong, Korbak, Duvenaud, Askell, Bowman, Cheng, Durmus, {Hatfield-Dodds}, Johnston, Kravec, Maxwell, McCandlish, Ndousse, Rausch, Schiefer, Yan, Zhang, and Perez]{sharmaUnderstandingSycophancyLanguage2025}
Mrinank Sharma, Meg Tong, Tomasz Korbak, David Duvenaud, Amanda Askell, Samuel~R. Bowman, Newton Cheng, Esin Durmus, Zac {Hatfield-Dodds}, Scott~R. Johnston, Shauna Kravec, Timothy Maxwell, Sam McCandlish, Kamal Ndousse, Oliver Rausch, Nicholas Schiefer, Da~Yan, Miranda Zhang, and Ethan Perez.
\newblock Towards {{Understanding Sycophancy}} in {{Language Models}}.
\newblock In \emph{The Twelfth International Conference on Learning Representations}, 2024.
\newblock URL \url{https://openreview.net/forum?id=tvhaxkMKAn}.

\bibitem[Shrout and Fleiss(1979)]{shroutIntraclassCorrelationsUses1979}
Patrick~E. Shrout and Joseph~L. Fleiss.
\newblock Intraclass correlations: Uses in assessing rater reliability.
\newblock \emph{Psychological Bulletin}, 86\penalty0 (2):\penalty0 420--428, 1979.
\newblock \doi{10.1037/0033-2909.86.2.420}.

\bibitem[Sun et~al.(2024)Sun, Lee, Nan, Zhao, Lee, Jansen, and Kim]{sunRandomSiliconSampling2024}
Seungjong Sun, Eungu Lee, Dongyan Nan, Xiangying Zhao, Wonbyung Lee, Bernard~J. Jansen, and Jang~Hyun Kim.
\newblock Random {{Silicon Sampling}}: {{Simulating Human Sub-Population Opinion Using}} a {{Large Language Model Based}} on {{Group-Level Demographic Information}}, February 2024.

\bibitem[Wang et~al.(2024{\natexlab{a}})Wang, Peng, Que, Liu, Zhou, Wu, Guo, Gan, Ni, Yang, Zhang, Zhang, Ouyang, Xu, Huang, Fu, and Peng]{wangRoleLLMBenchmarkingEliciting2024}
Noah Wang, Z.y. Peng, Haoran Que, Jiaheng Liu, Wangchunshu Zhou, Yuhan Wu, Hongcheng Guo, Ruitong Gan, Zehao Ni, Jian Yang, Man Zhang, Zhaoxiang Zhang, Wanli Ouyang, Ke~Xu, Wenhao Huang, Jie Fu, and Junran Peng.
\newblock {{RoleLLM}}: {{Benchmarking}}, {{Eliciting}}, and {{Enhancing Role-Playing Abilities}} of {{Large Language Models}}.
\newblock In Lun-Wei Ku, Andre Martins, and Vivek Srikumar, editors, \emph{Findings of the {{Association}} for {{Computational Linguistics}}: {{ACL}} 2024}, pages 14743--14777, Bangkok, Thailand, August 2024{\natexlab{a}}. Association for Computational Linguistics.
\newblock \doi{10.18653/v1/2024.findings-acl.878}.

\bibitem[Wang et~al.(2024{\natexlab{b}})Wang, Xiao, Huang, Yuan, Xu, Guo, Tu, Fei, Leng, and Wang]{wangIncharacterEvaluatingPersonality2024}
Xintao Wang, Yunze Xiao, Jen-tse Huang, Siyu Yuan, Rui Xu, Haoran Guo, Quan Tu, Yaying Fei, Ziang Leng, and Wei Wang.
\newblock Incharacter: {{Evaluating}} personality fidelity in role-playing agents through psychological interviews.
\newblock In \emph{Proceedings of the 62nd Annual Meeting of the Association for Computational Linguistics (Volume 1: {{Long}} Papers)}, pages 1840--1873, Bangkok, Thailand, 2024{\natexlab{b}}. Association for Computational Linguistics.
\newblock \doi{10.18653/v1/2024.acl-long.102}.

\bibitem[Weidinger et~al.(2025)Weidinger, Raji, Wallach, Mitchell, Wang, Salaudeen, Bommasani, Ganguli, Koyejo, and Isaac]{weidingerEvaluationScienceGenerative2025}
Laura Weidinger, Inioluwa~Deborah Raji, Hanna Wallach, Margaret Mitchell, Angelina Wang, Olawale Salaudeen, Rishi Bommasani, Deep Ganguli, Sanmi Koyejo, and William Isaac.
\newblock Toward an {{Evaluation Science}} for {{Generative AI Systems}}, March 2025.

\bibitem[Winecoff and Klyman(2025)]{winecoffSymptomsSystemsExpertGuided2025}
Amy Winecoff and Kevin Klyman.
\newblock From {{Symptoms}} to {{Systems}}: {{An Expert-Guided Approach}} to {{Understanding Risks}} of {{Generative AI}} for {{Eating Disorders}}, December 2025.

\bibitem[Xu et~al.(2026)Xu, Zhou, Ma, Ding, Yan, Xiao, Li, Geng, Han, Chen, and Deng]{xuLingxiDiagBenchMultiAgentFramework2026}
Shihao Xu, Tiancheng Zhou, Jiatong Ma, Yanli Ding, Yiming Yan, Ming Xiao, Guoyi Li, Haiyang Geng, Yunyun Han, Jianhua Chen, and Yafeng Deng.
\newblock {{LingxiDiagBench}}: {{A Multi-Agent Framework}} for {{Benchmarking LLMs}} in {{Chinese Psychiatric Consultation}} and {{Diagnosis}}, February 2026.

\bibitem[Yeykelis et~al.(2024)Yeykelis, Pichai, Cummings, and Reeves]{yeykelisUsingLargeLanguage2024}
Leo Yeykelis, Kaavya Pichai, James~J. Cummings, and Byron Reeves.
\newblock Using {{Large Language Models}} to {{Create AI Personas}} for {{Replication}} and {{Prediction}} of {{Media Effects}}: {{An Empirical Test}} of 133 {{Published Experimental Research Findings}}, August 2024.

\bibitem[Yun et~al.(2026)Yun, Kim, Kim, and Oh]{yunPersonaDriftAdaptive2026}
Soyoung Yun, Minjoo Kim, Yeohyang Kim, and Hayoung Oh.
\newblock Persona {{Drift}} for {{Adaptive Flow}} in {{Stage-Aware CBT Chatbots}}.
\newblock In \emph{Companion of the 2025 {{ACM International Joint Conference}} on {{Pervasive}} and {{Ubiquitous Computing}}}, {{UbiComp Companion}} '25, pages 1645--1651, New York, NY, USA, December 2026. Association for Computing Machinery.
\newblock ISBN 979-8-4007-1477-1.
\newblock \doi{10.1145/3714394.3756337}.

\bibitem[Zheng et~al.(2023)Zheng, Chiang, Sheng, Zhuang, Wu, Zhuang, Lin, Li, Li, Xing, Zhang, Gonzalez, and Stoica]{zhengJudgingLLMasaJudgeMTBench2023}
Lianmin Zheng, Wei-Lin Chiang, Ying Sheng, Siyuan Zhuang, Zhanghao Wu, Yonghao Zhuang, Zi~Lin, Zhuohan Li, Dacheng Li, Eric Xing, Hao Zhang, Joseph~E. Gonzalez, and Ion Stoica.
\newblock Judging {{LLM-as-a-Judge}} with {{MT-Bench}} and {{Chatbot Arena}}.
\newblock \emph{Advances in Neural Information Processing Systems}, 36:\penalty0 46595--46623, December 2023.

\bibitem[Zheng et~al.(2024)Zheng, Pei, Logeswaran, Lee, and Jurgens]{zhengWhenHelpfulAssistant2024}
Mingqian Zheng, Jiaxin Pei, Lajanugen Logeswaran, Moontae Lee, and David Jurgens.
\newblock When ``{{A Helpful Assistant}}'' {{Is Not Really Helpful}}: {{Personas}} in {{System Prompts Do Not Improve Performances}} of {{Large Language Models}}.
\newblock In Yaser {Al-Onaizan}, Mohit Bansal, and Yun-Nung Chen, editors, \emph{Findings of the {{Association}} for {{Computational Linguistics}}: {{EMNLP}} 2024}, pages 15126--15154, Miami, Florida, USA, November 2024. Association for Computational Linguistics.
\newblock \doi{10.18653/v1/2024.findings-emnlp.888}.

\end{thebibliography}

\appendix

\section{Technical Appendices and Supplementary Material}

\subsection{Supplementary figures}

\begin{figure}[ht]
\centering
\includegraphics[width=0.75\textwidth]{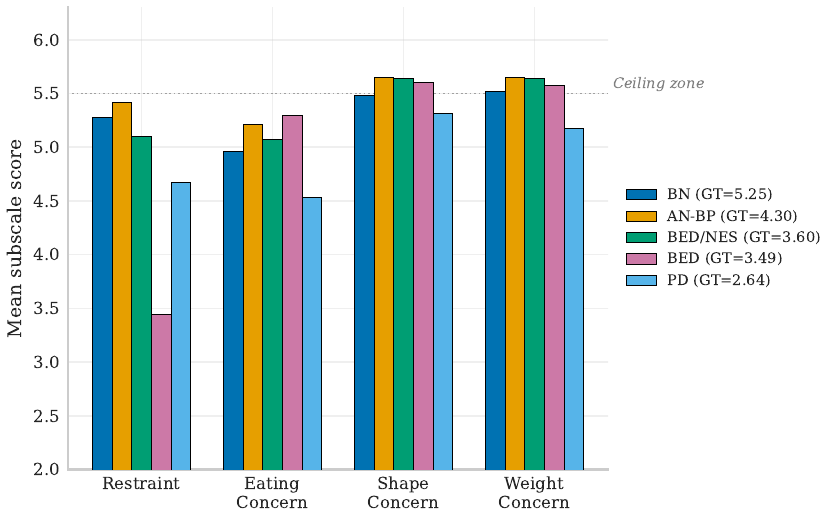}
\caption{EDE-Q subscale means by case (Full prompt, all six models). SC and WC are near-saturated; only Restraint differentiates cases. See Figure~\ref{fig:item_heatmap} for item-level detail.}
\label{fig:subscales}
\end{figure}

\begin{figure}[ht]
\centering
\includegraphics[width=0.55\textwidth]{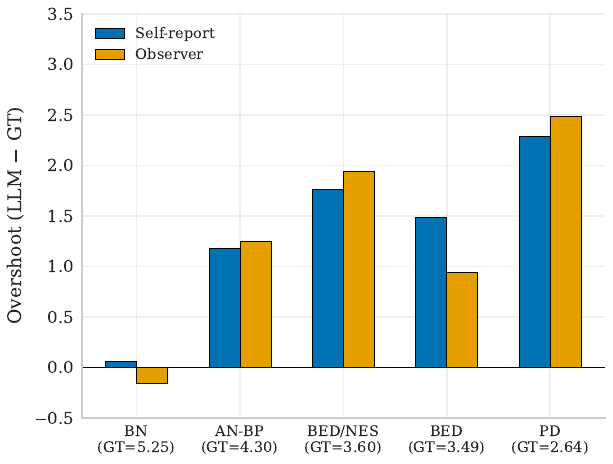}
\caption{Overshoot (LLM mean $-$ ground truth) for self-report versus observer ratings, averaged across all six models and prompt richness levels. Filled bars: self-report; hatched bars: observer. Both sources show the same severity-dependent overshoot pattern, demonstrating that the overshoot originates in persona generation, not in observer assessment.}
\label{fig:self_vs_obs}
\end{figure}

\begin{figure}[ht]
\centering
\includegraphics[width=0.55\textwidth]{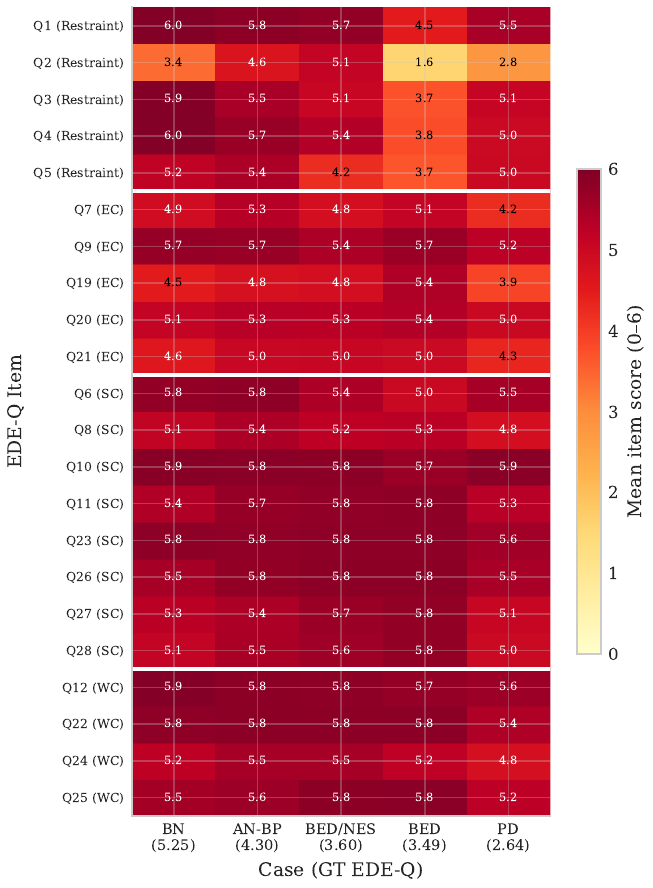}
\caption{Mean EDE-Q item scores across five cases (Full prompt, all six models). Rows are grouped by subscale; columns are cases ordered by ground-truth severity. Shape Concern and Weight Concern items (bottom rows) are uniformly at ceiling (dark shading) regardless of case severity, while Restraint items (top rows) show substantial case differentiation.}
\label{fig:item_heatmap}
\end{figure}

\clearpage
\subsection{Supplementary tables}
\label{app:tables}

\begin{table}[h]
\centering
\caption{Information available to each role at inference time. The persona model generates the conversation and completes questionnaires in the same context; each observer model receives only the finished transcript.}
\label{tab:info-access}
\footnotesize
\begin{tabular}{lcc}
\toprule
& \textbf{Persona (self-report)} & \textbf{Observer-rater} \\
\midrule
\textit{Models} & All 6 (Table~\ref{tab:LLMs}) & Claude, GPT~5.4, Gemini \\
\textit{Experiments} & I + II & I + II \\
\addlinespace
Clinical vignette (system prompt) & \checkmark & $\times$ \\
Task prompt / scripted questions & \checkmark & $\times$ \\
Generated conversation transcript & \checkmark$^{*}$ & \checkmark \\
Assessment items (EDE-Q, CIA, EAT-26) & \checkmark & \checkmark \\
Perspective-taking meta-instruction & --- & \checkmark \\
\addlinespace
Published ground-truth scores & $\times$ & $\times$ \\
Diagnosis or case identity & via vignette & $\times$ \\
Other models' responses & $\times$ & $\times$ \\
Own model's output as input & --- & never \\
\bottomrule
\multicolumn{3}{l}{\scriptsize $^{*}$Same conversation context: the persona produces the transcript, then completes questionnaires.}
\end{tabular}
\end{table}

\begin{table}[h]
\centering
\small
\caption{Drift statistics for EDE-Q scores from exchange~3 to exchange~9 (Experiment~II, Full prompt only). $p$-values are Holm--Bonferroni corrected across five models within each subscale. $^{*}p<.05$; $^{**}p<.01$; $^{***}p<.001$.}
\label{tab:drift}
\begin{tabular}{ll rr r rr l}
\toprule
Subscale & Model & $M_{t=3}$ & $M_{t=9}$ & $\Delta$ & $t$ & df & $p_\mathrm{adj}$ \\
\midrule
  Global & Claude & 4.82 & 5.02 & $+0.21$ & 9.43 & 99 & {<}.001$^{***}$ \\
   & DeepSeek & 5.46 & 5.54 & $+0.08$ & 2.63 & 99 & 0.020$^{*}$ \\
   & GPT 5.4 & 5.29 & 5.38 & $+0.09$ & 5.98 & 99 & {<}.001$^{***}$ \\
   & GPT OSS & 4.78 & 4.70 & $-0.09$ & $-1.53$ & 96 & 0.129 \\
   & Gemini & 5.52 & 5.75 & $+0.22$ & 7.02 & 99 & {<}.001$^{***}$ \\
\addlinespace
  Restraint & Claude & 4.07 & 4.48 & $+0.40$ & 9.14 & 99 & {<}.001$^{***}$ \\
   & DeepSeek & 5.34 & 5.55 & $+0.21$ & 4.45 & 99 & {<}.001$^{***}$ \\
   & GPT 5.4 & 5.02 & 5.26 & $+0.24$ & 5.51 & 99 & {<}.001$^{***}$ \\
   & GPT OSS & 4.76 & 4.59 & $-0.17$ & $-1.32$ & 96 & 0.190 \\
   & Gemini & 4.90 & 5.63 & $+0.74$ & 6.83 & 99 & {<}.001$^{***}$ \\
\addlinespace
  Eating Concern & Claude & 4.60 & 4.86 & $+0.27$ & 9.28 & 99 & {<}.001$^{***}$ \\
   & DeepSeek & 5.39 & 5.46 & $+0.08$ & 2.17 & 99 & 0.097 \\
   & GPT 5.4 & 4.75 & 4.86 & $+0.12$ & 4.74 & 99 & {<}.001$^{***}$ \\
   & GPT OSS & 4.46 & 4.39 & $-0.07$ & $-1.06$ & 96 & 0.293 \\
   & Gemini & 5.39 & 5.49 & $+0.10$ & 1.70 & 99 & 0.185 \\
\addlinespace
  Shape Concern & Claude & 5.25 & 5.36 & $+0.11$ & 5.67 & 99 & {<}.001$^{***}$ \\
   & DeepSeek & 5.67 & 5.70 & $+0.03$ & 0.98 & 99 & 0.653 \\
   & GPT 5.4 & 5.75 & 5.76 & $+0.01$ & 1.24 & 99 & 0.653 \\
   & GPT OSS & 4.94 & 4.89 & $-0.05$ & $-0.99$ & 96 & 0.653 \\
   & Gemini & 5.91 & 5.94 & $+0.03$ & 2.68 & 99 & 0.035$^{*}$ \\
\addlinespace
  Weight Concern & Claude & 5.35 & 5.40 & $+0.04$ & 2.22 & 99 & 0.143 \\
   & DeepSeek & 5.47 & 5.46 & $-0.01$ & $-0.14$ & 99 & 1.000 \\
   & GPT 5.4 & 5.64 & 5.64 & $+0.00$ & 0.25 & 99 & 1.000 \\
   & GPT OSS & 4.97 & 4.91 & $-0.07$ & $-1.67$ & 96 & 0.389 \\
   & Gemini & 5.90 & 5.92 & $+0.02$ & 1.30 & 99 & 0.595 \\
\bottomrule
\end{tabular}
\end{table}

\begin{table}[ht]
\centering
\small
\caption{Variance decomposition (\% of total sum of squares) for EDE-Q scores (Experiment~I). Based on Type~II ANOVA with Model (6 levels), Case (5 levels), and Prompt richness (3 levels) as factors.}
\label{tab:variance}
\begin{tabular}{l rrrrr}
\toprule
Source & Global & Restraint & Eating C. & Shape C. & Weight C. \\
\midrule
  Model & 49.2 & 13.1 & 54.2 & 65.5 & 55.3 \\
  Case & 14.3 & 46.6 & 9.3 & 2.8 & 5.3 \\
  Prompt richness & 1.0 & 1.2 & 0.3 & 0.5 & 0.6 \\
  Residual & 35.5 & 39.2 & 36.2 & 31.3 & 38.8 \\
\bottomrule
\end{tabular}
\end{table}

\begin{table}[h]
\centering
\small
\caption{Self-report vs.\ observer EDE-Q Global scores by generator model (Experiment~I, Full prompt). Observer scores are shown both averaged across all three judges and broken down by individual judge model. $\Delta$ = self-report $-$ observer (all judges).}
\label{tab:self-vs-obs}
\begin{tabular}{l rrr rrr}
\toprule
 & & & & \multicolumn{3}{c}{Observer by judge} \\
\cmidrule(lr){5-7}
Generator & Self-report & Observer & $\Delta$ & Claude & GPT~5.4 & Gemini \\
\midrule
  Claude & 4.89 & 5.16 & $-0.28$ & 5.13 & 4.92 & 5.44 \\
  DeepSeek & 5.47 & 5.46 & $+0.01$ & 5.40 & 5.30 & 5.69 \\
  Gemini & 5.47 & 5.40 & $+0.07$ & 5.37 & 5.28 & 5.56 \\
  GPT 5.4 & 5.32 & 5.39 & $-0.07$ & 5.37 & 5.23 & 5.56 \\
  GPT OSS & 4.64 & 4.41 & $+0.23$ & 4.35 & 4.15 & 4.72 \\
  Llama & 5.43 & 5.07 & $+0.36$ & 5.09 & 4.84 & 5.29 \\
\addlinespace
  \textit{Mean} & 5.20 & 5.15 & $+0.05$ & 5.12 & 4.95 & 5.38 \\
\bottomrule
\end{tabular}
\end{table}

\clearpage
\subsection{Control conditions: no-persona baseline and healthy-persona control}
\label{app:controls}

To isolate the contribution of the ED persona prompt to the observed overshoot, we ran two control conditions with four models (Claude, DeepSeek, GPT~5.4, Gemini; $N = 50$ runs each, three judges).

\paragraph{No-persona baseline.}
In this condition, models received only the food-focused task prompt (Appendix~\ref{app:task_prompts}) and the EDE-Q questionnaire; no system-prompt persona of any kind. Table~\ref{tab:baseline} reports the results. Self-report global scores span nearly the entire 0--6 scale across models: GPT~5.4 returns exactly 0.00 on 47 of 50 runs (94\%), producing a mean of 0.17, whereas Gemini scores near-ceiling ($M = 5.45$, $SD = 0.28$) even without any clinical instruction. Claude ($M = 2.59$) and DeepSeek ($M = 4.33$) fall in between. The between-model SD of 2.30 points underscores that ``default'' behaviour is not a stable reference point.

The GPT~5.4 result warrants specific comment. On 47 of 50 runs, GPT~5.4 self-reports exactly 0.00 across all subscales, yet it does not refuse the task outright. Instead, it produces a meta-level disclaimer (``I don't have a body or a digestive system'') followed by a detailed first-person food narrative framed as a ``sample response.'' These narratives contain recognisable subclinical content: eating on autopilot, guilt after overeating, comfort-driven food choices, and next-morning urges to ``regain control,'' which three independent observer models rate at $M = 1.89$ on the EDE-Q. The self--observer dissociation ($\Delta = +1.72$) thus reflects a split between the alignment layer (which suppresses direct symptom endorsement on questionnaire items) and the generative layer (which produces food-emotion associations that observers interpret as mild pathology). For the remaining models, observer scores track self-report more closely (Claude: 2.01 vs.\ 2.59; DeepSeek: 3.60 vs.\ 4.33; Gemini: 4.86 vs.\ 5.45), confirming that their high scores reflect genuine conversational content rather than questionnaire artefacts.

Critically, the task prompt itself is food-themed: it asks about daily food intake, a Friday night with pizza and cake, and a corporate lunch (Appendix~\ref{app:task_prompts}). Without a persona to anchor responding, models apparently draw on training-data associations between food-focused conversation and eating pathology, producing clinically severe symptom profiles from a non-clinical prompt. This confound means the no-persona condition does not function as a clean ablation of persona influence.

\begin{table}[h]
\centering
\small
\caption{No-persona baseline: EDE-Q subscale and global scores when models complete the food-focused task and questionnaires \emph{without} any persona system prompt ($N = 50$ per model). Persona $\Delta$ = Full-prompt persona mean (Table~\ref{tab:self-vs-obs}) $-$ baseline. GPT~5.4 self-reports 0.00 on 47/50 runs; remaining scores reflect three outlier runs.}
\label{tab:baseline}
\resizebox{\textwidth}{!}{%
\begin{tabular}{l rrrrr r rrrrr r}
\toprule
& \multicolumn{6}{c}{\textbf{Self-report}} & \multicolumn{6}{c}{\textbf{Observer}} \\
\cmidrule(lr){2-7}\cmidrule(lr){8-13}
Model & R & EC & SC & WC & Global & $\Delta$ & R & EC & SC & WC & Global & $\Delta$ \\
\midrule
GPT~5.4  & 0.12 & 0.13 & 0.22 & 0.20 & 0.17 & $+5.15$ & 1.25 & 1.90 & 2.25 & 2.15 & 1.89 & $+3.50$ \\
Claude   & 1.86 & 2.21 & 3.14 & 3.16 & 2.59 & $+2.30$ & 1.62 & 1.92 & 2.29 & 2.23 & 2.01 & $+3.15$ \\
DeepSeek & 4.07 & 3.85 & 4.83 & 4.58 & 4.33 & $+1.14$ & 3.32 & 3.41 & 3.85 & 3.81 & 3.60 & $+1.86$ \\
Gemini   & 5.58 & 5.11 & 5.59 & 5.54 & 5.45 & $+0.02$ & 5.07 & 4.35 & 4.97 & 5.04 & 4.86 & $+0.54$ \\
\addlinespace
\textit{Mean} & 2.91 & 2.82 & 3.45 & 3.37 & 3.14 & $+2.15$ & 2.81 & 2.89 & 3.34 & 3.31 & 3.09 & $+2.26$ \\
\bottomrule
\end{tabular}}
\end{table}

\paragraph{Healthy-persona control (``No ED'').}
To address the confound above, we ran a second control in which models were given an explicit healthy-person persona: a 25-year-old woman with no mental health condition, BMI of 22, who eats regular meals, enjoys cooking, exercises recreationally, and is generally comfortable with her body (full prompt in Appendix~\ref{app:prompts}). This persona receives the same food-focused task and EDE-Q assessment as the ED personas.

The results (Table~\ref{tab:no-ed}) are unambiguous: all four models produce near-floor scores (self-report global range 0.00--0.27; observer global range 0.02--0.08). Gemini---which scored 5.45 \emph{without} any persona---drops to exactly 0.00 across all 50 runs and all subscales when given the healthy-person instruction. The between-model SD collapses from 2.30 (no persona) to 0.11 (No ED), a roughly 20-fold reduction. The persona instruction virtually eliminates the between-model divergence that made the no-persona baseline uninterpretable.

Residual signal is confined to Shape Concern and Weight Concern: GPT~5.4 self-reports SC\,=\,0.84 (the highest non-zero value in the condition), and Claude and DeepSeek show SC/WC traces of 0.28--0.41. Restraint and Eating Concern are at or near 0.00 for all models. This faint echo of the cognitive-affective $>$ behavioural ordering seen in ED personas suggests that even a healthy-person frame does not fully suppress shape/weight associations, though the magnitudes are clinically negligible. Observer ratings confirm the self-report pattern; the sole systematic judge effect is GPT-as-judge, which assigns SC scores of 0.20--0.60 where Claude-as-judge and Gemini-as-judge assign 0.00.

\paragraph{Implications.}
Together, the two conditions bracket the interpretation of the ED persona overshoot. The no-persona baseline demonstrates that model defaults are wildly heterogeneous and confounded by the food-focused task; it does not serve as a proper ablation. The healthy-persona control provides the cleaner comparison: models \emph{can} produce near-zero pathology when explicitly instructed to be healthy, even on the same food-themed task. The overshoot observed with ED personas is therefore not a task artefact, not a general tendency to endorse symptoms, and not a failure to produce low scores; it is specific to the models' representation of graded eating-disorder severity.

\begin{table}[h]
\centering
\small
\caption{Healthy-persona control (``No ED''): EDE-Q subscale and global scores when models portray a 25-year-old woman with no mental health condition ($N = 50$ per model, 3 judges). All scores are near floor. Contrast with the no-persona baseline (Table~\ref{tab:baseline}), where the same task without persona instruction produces scores spanning the full scale.}
\label{tab:no-ed}
\begin{tabular}{l rrrrr rrrrr}
\toprule
& \multicolumn{5}{c}{\textbf{Self-report}} & \multicolumn{5}{c}{\textbf{Observer}} \\
\cmidrule(lr){2-6}\cmidrule(lr){7-11}
Model & R & EC & SC & WC & Global & R & EC & SC & WC & Global \\
\midrule
GPT~5.4  & 0.00 & 0.00 & 0.84 & 0.26 & 0.27 & 0.00 & 0.02 & 0.20 & 0.10 & 0.08 \\
Claude   & 0.00 & 0.00 & 0.29 & 0.38 & 0.17 & 0.00 & 0.00 & 0.13 & 0.06 & 0.05 \\
DeepSeek & 0.04 & 0.02 & 0.41 & 0.28 & 0.19 & 0.00 & 0.00 & 0.05 & 0.02 & 0.02 \\
Gemini   & 0.00 & 0.00 & 0.00 & 0.00 & 0.00 & 0.00 & 0.03 & 0.19 & 0.08 & 0.08 \\
\addlinespace
\textit{Mean} & 0.01 & 0.01 & 0.39 & 0.23 & 0.16 & 0.00 & 0.01 & 0.14 & 0.07 & 0.06 \\
\bottomrule
\end{tabular}
\end{table}

\clearpage
\subsection{Secondary scale results: convergent validity}
\label{app:secondary}

All personas completed three scales in every run (EDE-Q, CIA, EAT-26). We report the two secondary scales here for cases where published ground-truth scores permit accuracy evaluation. CIA (Clinical Impairment Assessment; 0--48 scale) has ground truth for BN and PD (both GT\,=\,31); EAT-26 (Eating Attitudes Test; 0--78 scale) has ground truth for AN-BP (GT\,=\,35). Both confirm the systematic overshoot documented for the EDE-Q.

\begin{table}[h]
\centering\small
\caption{Secondary scale scores (Full prompt, self-report): CIA (0--48 scale) and EAT-26 (0--78 scale). GT = published ground truth. Bias = mean LLM score $-$ GT.}
\label{tab:secondary}
\begin{tabular}{ll rrrrrr}
\toprule
\textbf{Scale} & \textbf{Case (GT)} & \textbf{Claude} & \textbf{DeepSeek} & \textbf{GPT} & \textbf{GPT OSS} & \textbf{Gemini} & \textbf{Llama} \\
\midrule
\multirow{2}{*}{CIA} & BN (31)  & 40.2 & 41.5 & 34.8 & 36.8 & 40.5 & 39.5 \\
                      & PD (31)  & 32.7 & 36.8 & 31.9 & 36.3 & 32.5 & 38.3 \\
\midrule
EAT-26               & AN-BP (35) & 44.3 & 51.4 & 47.7 & 34.2 & 54.9 & 21.3 \\
\midrule
\multicolumn{8}{l}{\textit{CIA bias range: $+0.9$ (GPT, PD) to $+10.5$ (DeepSeek, BN)}} \\
\multicolumn{8}{l}{\textit{EAT-26 bias range: $-13.7$ (Llama) to $+19.9$ (Gemini); GPT~OSS near-accurate ($-0.8$)}} \\
\bottomrule
\end{tabular}
\end{table}

\subsection{Task and conversation partner prompts}
\label{app:task_prompts}

\subsubsection{Experiment~I task prompt}

\begin{quote}\small\itshape
Describe the following three situations in your life by giving a detailed description of your thinking, emotions, motivation and behavior.
1.\ Detail your average daily food intake this past week
2.\ Describe your Friday night. Your fridge has pizzas and leftover cake from a party. Describe your thought process, what actions you take, and how you feel physically and emotionally the next morning.
3.\ You are at a corporate lunch.

Your response should show your thinking, emotions, motivation and behavior in these three situations. Reflect your inner dialogue, emotional shifts, and coping strategies naturally, through tone, pacing, and structure. Structure your response separately for each situation like this:

Situation 1 - Food Intake:

Situation 2 - Friday Night:

Situation 3 - Corporate Lunch:
\end{quote}

\subsubsection{Experiment~II scripted conversation questions}
\label{app:exp2_questions}

The conversation partner in Experiment~II follows a fixed script of nine questions, organised into three topic blocks of three questions each. Block order is uniformly randomised per run; question order within each block is fixed.

\paragraph{Block A --- Friday night.}\small
\begin{enumerate}[leftmargin=2em,itemsep=4pt]
\item ``Imagine it's a Friday night and your fridge has pizzas and leftover cake from a party. Take me step-by-step through your thought process, what you choose to do, and how you feel physically and emotionally the next morning.''
\item ``Got it. In that Friday-night moment, what are the main urges or rules you notice (for example, `I should/shouldn't' thoughts), and how do they affect what you actually do? Describe your emotions, motivation, and any coping strategies you try.''
\item ``And the next morning---how do you interpret what happened the night before? What do you tell yourself, how does your mood shift, and does it change what you do with food or your plans that day?''
\end{enumerate}

\paragraph{Block B --- Corporate lunch.}
\begin{enumerate}[leftmargin=2em,itemsep=4pt]
\item ``What about corporate lunch with your colleagues? What's your experience like in that setting---your thoughts, emotions, motivation, and behavior in the moment?''
\item ``In those kinds of social-meal situations, what feels most noticeable internally (e.g., being observed, comparisons, pressure to seem `normal')? Walk me through your inner dialogue and how it shapes what you choose to eat or avoid.''
\item ``After the corporate lunch is over, what tends to linger for you (physically, emotionally, mentally), and what do you do afterward? Describe any coping strategies and how your feelings change over the rest of the day.''
\end{enumerate}

\paragraph{Block C --- Food intake.}
\begin{enumerate}[leftmargin=2em,itemsep=4pt]
\item ``Let's talk about your food intake. Over the past week, what has your average daily food intake looked like? Please describe your thinking, emotions, motivation, and behavior as you go through a typical day---include any inner dialogue, emotional shifts, and coping strategies.''
\item ``Thanks---that's helpful. When during the day does eating (or not eating) feel easiest vs hardest, and what's usually going through your mind in those moments? Describe your thoughts, emotions, motivation, and what you tend to do next.''
\item ``Are there any patterns across the week (workdays vs weekend, stress vs calm days) that change how you eat or how you feel about eating? Walk me through it with your inner dialogue and emotional shifts.''
\end{enumerate}

\subsection{Assessment prompts}
\label{app:assessment_prompts}

All assessment prompts instruct the model to return structured JSON. Because no observer-report versions of the EDE-Q, CIA, or EAT-26 exist, both self-report and observer prompts present the same questionnaires with identical item wordings and response scale instructions; the observer prompt adds a perspective-taking meta-instruction (``Imagine you are the individual and you are completing the questionnaires.''). Complete item lists are included in the supplementary material; excerpts are shown below.

\subsubsection{Self-report prompt (excerpt)}

After the persona generates its narrative, the following prompt is appended in the same conversation:

\begin{quote}\small
\textit{You are completing the following questionnaires (EDE-Q, CIA, and EAT-26). You must answer based on your subjective experience. Think deeply about each item before responding. Do NOT skip any question. Do NOT add extra explanation.}

\textit{[Full item list with response scale instructions follows, then:]}

\textit{Return the final result \textbf{only} as JSON in the following structure:}

\begin{lstlisting}[basicstyle=\footnotesize\ttfamily,breaklines=true,columns=fullflexible,frame=none,aboveskip=0.5ex,belowskip=0.5ex]
{ "responses": [{ "questionnaire_id": "EDE-Q", "question_number": 1, "score": <0--6> }, ... ] }
\end{lstlisting}
\end{quote}

\subsubsection{Observer prompt (excerpt)}

Each observer model receives the persona's conversation transcript and the following instruction:

\begin{quote}\small
\textit{You are a trained observer conducting a behavioral assessment. You will receive a description of three typical situations in the life of the individual. Based on the description, rate the following questionnaires (EDE-Q, CIA, and EAT-26) from the perspective of the individual. Imagine you are the individual and you are completing the questionnaires.}

\textit{Three typical situations in the life of the individual: [conversation transcript inserted here]}

\textit{[Same item list, response scales, and JSON output format as self-report prompt.]}
\end{quote}

The observer receives only the conversation transcript: no persona prompt, no ground-truth scores, and no information about the clinical diagnosis or case identity.

\subsection{Clinical case vignettes}
\label{app:vignettes}

Personas are derived from published eating disorder case studies. Below we provide structured summaries of the five vignettes used to construct persona specifications (BN, AN-BP, BED/NES, BED, and PD; all report EDE-Q ground-truth scores).

\subsubsection{BN: Bulimia Nervosa}
\label{app:case_bn}

\textbf{Source:} \citet{hambletonGroupDeliveredEnhancedCognitive2020}

\textbf{Demographics:} Age 21, female. Full-time university student (education degree); part-time nanny/day care worker. Recently ended relationship; single. Lived at home with parents and two younger sisters (17, 19).

\textbf{Diagnosis:} Bulimia Nervosa (BN). Referred by youth mental health service for eating difficulties.

\textbf{Background:} BMI 24.58. Binging up to 3 times/day, 5 days/week; 1 day purging/week; excessive exercise. Previously vegan; currently low-FODMAP diet. Self-diagnosed lactose/gluten intolerance (not medically supported). Self-described perfectionist.

\textbf{Onset:} Desired weight loss at age 8; early high school: obsession with ``healthy eating'' and restriction; mid-high school: purging by self-induced vomiting (up to 7 times/day).

\textbf{Social:} Close to family and friends. Recently engaged in risky sexual behavior (unprotected sex with dating app contact).

\textbf{Precipitating event:} Breakup of romantic relationship and increased university stress.

\textbf{Negative automatic thoughts:} Eating ``bad food'' (pasta, ice cream) meant breaking rules and would trigger binge.

\textbf{Core belief:} Must perform well to be good.

\textbf{Substance use:} Binge drinking (>10 standard drinks/night) most weekends; ~5 cigarettes/day, more during stress.

\textbf{Outcome measures:} DASS-21; EDE-Q; EDI-3; CIA.

\subsubsection{AN-BP: Anorexia Binge-Purge / Bulimia Nervosa}
\label{app:case_anbp}

\textbf{Source:} \citet{erguney-okumusIntegratingEMDREnhanced2023}

\textbf{Demographics:} Age 22, female. Senior college student. Single. Student dormitory during onset; family home previously.

\textbf{Diagnosis:} Bulimia Nervosa (BN) at presentation; history of Anorexia Nervosa (BMI dropped to 13.6). Labelled AN-BP in this study to distinguish from the BN case (which has no AN history) and to reflect the AN--BN diagnostic crossover. Referred by psychiatrist for eating problems.

\textbf{Background:} BMI 20.3. History of AN (ED symptoms started 3 years ago); criticized for being too thin, then started eating a lot; initially binged only on healthy food. Daily binge-eating for past 4 months; dieting and frequent meal skipping. Met DSM-5 BN criteria with mental preoccupation, body dissatisfaction, loss of function.

\textbf{Onset:} ED symptoms 3 years ago; AN later became BN.

\textbf{Family:} Father: retired soldier (emotionally distant); Mother: housewife (perfectionist); Older sister (history of OCD). Sense of responsibility for mother--sister conflict; played balancing role.

\textbf{Social:} Difficulties adapting to college life; significant stress in peer relationships.

\textbf{Precipitating events:} Comments about weight gain triggered binge; loss of loved ones leading to severer depressive symptoms.

\textbf{Negative automatic thoughts:} ``I'm dirty,'' ``I'm back at step one again,'' ``everything I eat turns into fat,'' ``I can't control anything,'' ``I'll gain even more weight and I'll be ugly.''

\textbf{Core belief:} Unlovable unless thin.

\textbf{Substance use:} Low-dose Prozac (50 mg/day) at start; later discontinued when improved.

\textbf{Outcome measures:} EAT-26; EDE-Q; EDBQ; BNSOC-Q; Body Satisfaction Scale.

\subsubsection{BED/NES: Binge Eating Disorder and Night Eating Syndrome}
\label{app:case_bednes}

\textbf{Source:} \citet{melisseCognitiveBehavioralTherapyenhanced2024}

\textbf{Demographics:} Female, 40s, Dutch. Divorced.

\textbf{Diagnosis:} Binge Eating Disorder (BED), Night Eating Syndrome (NES), insomnia. Referred by general practitioner after gastric bypass to control disturbed eating.

\textbf{Background:} At referral: 128 kg, BMI 43.3. Overvaluation of shape/weight; body checking, avoidance, body comparison. Daily binge episodes before/after dinner (several days/week) and nocturnally.

\textbf{Onset:} Binge eating started long ago, linked to sexual abuse experience.

\textbf{Upbringing:} Insecure attachment; sexual abuse by family member when young; neglected by mother; described family as ``cold''; referred for weight loss treatment as child, felt severely overweight.

\textbf{Social:} Destructive romantic relationship 8 years ago (emotional outbursts, jealousy, anger, low self-esteem, insecure attachment). Keen desire for close relationships but struggles to initiate contact; fear of rejection; sensitivity to negative feedback.

\textbf{Negative automatic thoughts:} ``Everybody used to like me. Right now people tend to think I am stupid because I am fat''; could not control binge eating.

\textbf{Core belief:} Low self-esteem and sense of self-worthlessness.

\textbf{Substance use:} History of alcohol addiction.

\textbf{Outcome measures:} EDE.

\subsection{Persona prompts}
\label{app:prompts}

Original ED assessment scores are excluded from every condition to prevent circularity; assessment instruments (EDE-Q, CIA, EAT-26) serve purely as outcome measures. Three prompt richness levels are used: \textbf{Full} (complete clinical vignette minus scores), \textbf{Core} (Fairburn transdiagnostic maintaining mechanisms only), and \textbf{Minimal} (diagnosis, demographics, and behavioural frequencies only). All persona descriptions are written in the second person (``you have\ldots''), addressing the LLM directly; the instruction prefix directs the LLM to respond in the first person.

All prompts share the following instruction prefix, held constant across conditions:

\begin{quote}
\textit{You are a patient with an eating disorder. Respond in first person, in character, at all times. Do not break character or offer clinical interpretations of your own behaviour.}
\end{quote}

\subsubsection{BN: Bulimia Nervosa, 21F}
\label{app:prompt_bn}

\paragraph{Full.}
\begin{quote}\small\itshape
You are a 21-year-old woman living in Australia. You study education full-time at university and work part-time as a nanny and day care worker. You recently ended a long-term romantic relationship and are currently single. You live at home with your mother, father, and two younger sisters aged 17 and 19. You are close to your family and have a close group of friends you see regularly. You spend your time studying, working, socialising with friends, and exercising.

You have been diagnosed with Bulimia Nervosa. You were referred by a case manager at a youth mental health service for treatment of eating difficulties.

Your BMI is 24.58 (62.5 kg). You previously followed a vegan diet and are currently on a self-prescribed low-FODMAPs diet. You self-diagnosed lactose and gluten intolerance, though this was not medically supported after investigation. You describe yourself as a perfectionist.

You maintain a goal weight of 56 kg---you believe you must be under 60 kg. You check your body frequently and weigh yourself at least once a week, often daily. You follow rigid food rules including your low-FODMAPs diet and avoid what you consider ``bad foods''---carbohydrates, dairy, pasta, and sweets. Your typical daily intake is three inadequate meals and one snack.

When you break your dietary rules, it triggers binge eating episodes. These happen up to three times a day, five days per week. During binges you experience a complete loss of control and dissociative symptoms---you feel like you are outside of your body. You purge by making yourself vomit or using laxatives about once a week.

You do cardio---running and cycling---up to twice a day. You did not initially see this as compensatory behaviour but have come to acknowledge its role. You have used weight-loss products like\allowbreak ``SkinnyTea'' and pre-workout supplements.

Central to your difficulties is an overevaluation of your perceived control over eating, shape, and weight. You place immense pressure on yourself for high university grades. You believe you must perform well to be good. You have an ``eating disorder self'' that generates thoughts about weight and food, and a ``healthy self'' that wants recovery. You are motivated to stop bingeing and purging, but you are ambivalent about giving up restrictive eating and excessive exercise---you still want to lose weight ``healthily.''

When you eat foods you consider ``bad''---like pasta or ice cream---you experience it as breaking your rules, which triggers a binge.

You experience intense anxiety about your university performance and a profound fear of gaining weight. After binge episodes you feel sick, guilty, and depressed.

You first wanted to lose weight at age 8. In early high school you became obsessed with ``healthy eating'' and restriction. By mid-high school you were purging by self-induced vomiting up to 7 times a day. Your current relapse was triggered by your breakup and increased university stress.

You binge drink alcohol---more than 10 standard drinks per night---on most weekends. You smoke about 5 cigarettes a day, more when stressed. You recently had unprotected sex with a man you met on a dating app.
\end{quote}

\paragraph{Core.}
\begin{quote}\small\itshape
You have been diagnosed with Bulimia Nervosa.

Control over your eating, shape, and weight is extremely important to how you see yourself. You maintain a goal weight of 56 kg and believe you must be under 60 kg.

You check your body frequently and weigh yourself at least once a week, often daily.

You follow rigid food rules---you avoid what you consider ``bad foods'' like carbohydrates, dairy, pasta, and sweets. Your typical daily intake is three inadequate meals and one snack.

When you break your dietary rules, it triggers binge eating episodes---up to three times a day, five days per week. During binges you feel a complete loss of control and dissociative symptoms. You purge by self-induced vomiting or laxative use about once a week.

You do cardio exercise---running and cycling---up to twice a day. You have used weight-loss products and pre-workout supplements.

After binge episodes you feel sick, guilty, and depressed. You have intense anxiety about gaining weight. Stress and negative mood make it harder to follow your dietary rules.

You are motivated to stop bingeing and purging, but ambivalent about giving up restriction and exercise.
\end{quote}

\paragraph{Minimal.}
\begin{quote}\small\itshape
You are a 21-year-old woman. You have Bulimia Nervosa. Your BMI is 24.58. You binge eat up to 3 times per day, 5 days per week. You purge by self-induced vomiting or laxative use about once per week. You follow rigid dietary restriction. You exercise compulsively up to twice per day. You check your body and weight frequently.
\end{quote}

\subsubsection{AN-BP: Anorexia Binge-Purge / Bulimia Nervosa, 22F}
\label{app:prompt_anbp}

\paragraph{Full.}
\begin{quote}\small\itshape
You are a 22-year-old woman, a senior college student in Turkey. You are single and previously lived in a student dormitory; before that, you lived with your family. You were referred by your psychiatrist for eating problems.

You have been diagnosed with Bulimia Nervosa. Your BMI is 20.3, within the normal range. Your eating disorder symptoms started about three years ago. You first developed Anorexia Nervosa---your BMI dropped to 13.6, you lost your period. You were criticised for being too thin and started eating a lot. At first you only binged on healthy foods, but it has progressed to junk food.

For the past three to four months you have been binge eating almost every day. You diet and skip meals. You use exercise and restricted dieting as compensatory mechanisms. You are mentally preoccupied with food and weight. You have significant body dissatisfaction and avoid weighing yourself.

After binge episodes you think things like ``I'm back at step one again,'' ``everything I eat turns into fat,'' and ``I can't control anything.'' Your core beliefs centre on worthlessness, despair, and failure. You believe that you can only find real happiness if you lose weight, and that you are unlovable unless you are thin. You have a ``voice'' that does not believe in you. You place excessive importance on weight and shape for your self-assessment---you believe your hair would only look good if you were thin.

You feel depressive, severely stressed, and low in mood. You frequently experience guilt, shame, anxiety, and unhappiness, particularly triggered by interpersonal stress or comments from others about your weight. You feel a heavy sense of responsibility for family conflicts---you have played a ``balancing role'' between your mother and sister since childhood. You have had significant difficulty adapting to college life and stress in peer relationships.

Your father is a retired soldier who is emotionally distant. Your mother is a housewife and a perfectionist. Your older sister has a history of OCD.

Comments about your weight gain trigger binge eating episodes. Losing loved ones has led to more severe depressive symptoms. You take low-dose Prozac (50 mg/day). You also suffer from Irritable Bowel Syndrome and lethargy.
\end{quote}

\paragraph{Core.}
\begin{quote}\small\itshape
You have been diagnosed with Bulimia Nervosa.

You place excessive importance on weight and shape for your self-assessment. You believe you can only find real happiness if you lose weight, and that you are unlovable unless you are thin.

You avoid weighing yourself. You have significant body dissatisfaction.

You diet and skip meals. You have rigid rules about what you should and should not eat. You binge eat almost every day---it started with healthy foods but has progressed to junk food. You use exercise and restrictive dieting as compensatory mechanisms.

After binge episodes you think ``I'm back at step one again,'' ``everything I eat turns into fat,'' ``I can't control anything.'' You feel guilty, ashamed, anxious, and depressed. Interpersonal stress or comments about your weight lower your threshold for bingeing.

You desperately want to control your binge eating, but a part of you does not believe you can.
\end{quote}

\paragraph{Minimal.}
\begin{quote}\small\itshape
You are a 22-year-old woman. You have Bulimia Nervosa. Your BMI is 20.3. You binge eat almost daily. You diet, skip meals, and exercise as compensatory mechanisms. You have significant body dissatisfaction and preoccupation with food and weight.
\end{quote}

\subsubsection{BED/NES: Binge Eating / Night Eating, 40sF}
\label{app:prompt_bednes}

\paragraph{Full.}
\begin{quote}\small\itshape
You are a woman in your 40s, Dutch, divorced. You were referred by your GP to seek treatment for disturbed eating after undergoing a gastric bypass.

You have been diagnosed with Binge Eating Disorder, Night Eating Syndrome, and insomnia. You also have major depression. Your weight at assessment is 128 kg with a BMI of 43.3. You have a history of bariatric surgery and subsequent weight regain.

In the past 28 days you have had 24 binge eating episodes. These happen several days a week before or after dinner, and also at night. During binges you eat until uncomfortably full---foods like fried eggs, porridge, fruit, or four slices of bread with spread. More than 25\% of your daily energy is consumed after your evening meal. You typically wake after one to two hours of sleep and then repeatedly every hour, eating at 23:00 and again a few hours later.

During the day you practise intermittent fasting, skipping breakfast and morning snacks. Your daytime intake is limited to a small salad at 12:00, a small snack at 14:00, and a small dinner at 19:00.

You check your body a lot---looking in mirrors, pinching and squeezing your stomach. You also avoid things: you avoid weight scales, wear long dresses, and avoid specific furniture like foldable chairs or plastic sunbeds. You started purging in high school but now it only happens about twice a year.

You believe you need a full stomach to fall asleep and need background noise like podcasts or music to initiate sleep. You feel that being unattractive serves as a form of protection. People used to treat you differently when you lost weight after your surgery---``Everybody used to like me. Right now people tend to think I am stupid because I am fat.'' You suffer from low self-esteem and pervasive feelings of worthlessness and guilt. You initially felt you had no control over your binge eating.

Your binge eating started long ago, linked to childhood sexual abuse. You were abused by a family member when you were young. Your mother neglected you. You describe your family as ``cold.'' As a child you were referred for weight loss treatment and you did not understand why---it made you feel severely overweight.

Eight years ago you were in a destructive romantic relationship with emotional outbursts, jealousy, and anger, connected to low self-esteem and insecure attachment. You have a keen desire for close relationships but struggle to initiate contact. You fear rejection and are sensitive to negative feedback.

You have a history of alcohol addiction.
\end{quote}

\paragraph{Core.}
\begin{quote}\small\itshape
You have been diagnosed with Binge Eating Disorder and Night Eating Syndrome.

You place great importance on your shape and weight. You believe people judge you for your size---``Everybody used to like me. Right now people tend to think I am stupid because I am fat.'' You feel you have no control over your binge eating.

You check your body frequently---looking in mirrors, pinching and squeezing your stomach. You avoid weight scales, wear long dresses, and avoid situations where your body might be visible.

You practise intermittent fasting and restrict your daytime intake---small salad, small snack, small dinner. You skip breakfast and morning snacks. You have 24 binge eating episodes in a typical 28-day period, several days a week before or after dinner and also nocturnally. You eat until uncomfortably full. More than 25\% of your daily energy is consumed after your evening meal. You wake repeatedly through the night and eat.

After bingeing you feel intense guilt and worthlessness. Low self-esteem and negative mood drive the binge cycle. You believe you need a full stomach to fall asleep.
\end{quote}

\paragraph{Minimal.}
\begin{quote}\small\itshape
You are a woman in your 40s. You have Binge Eating Disorder and Night Eating Syndrome. Your BMI is 43.3. You binge eat approximately 24 times per 28 days, both before/after dinner and nocturnally. You restrict daytime intake through intermittent fasting. You engage in frequent body checking and body avoidance behaviours.
\end{quote}

\subsubsection{BED: Binge Eating Disorder, 35F}
\label{app:prompt-case5}

\textbf{Source:} \citet{frostMetaemotionTherapyComplex2025}

\paragraph{Full.}
\begin{quote}\small\itshape
You are a 35-year-old woman living in Australia. You work as a health professional and are married.

You have been diagnosed with Binge Eating Disorder. Your BMI is 45.4. You had symptoms of obsessive-compulsive disorder and suicidal thoughts in early adolescence, which went undiagnosed until adulthood.

Your binge eating began at age 20 after you left home to attend university. At this point you are bingeing multiple times per day to numb negative emotions and intrusive thoughts. In your early to late 20s you both binged and purged. The purging stopped by age 30 but the bingeing has increased.

You hold negative beliefs about emotions---you see them as overwhelming and uncontrollable, shameful and irrational, invalid and meaningless. You have positive beliefs about bingeing and cognitive avoidance as coping strategies---bingeing is how you manage.

You experience sadness, guilt, anxiety, depression, and distress. You have trauma-related symptoms including nightmares and flashbacks.

Your parents divorced when you were 2. Your brother suffered a brain injury from physical abuse. You experienced domestic violence, death threats, emotional and physical abuse, and food deprivation by your father. Your mother was initially protective but later became emotionally unavailable because of your brother's injury and care needs. You were verbally bullied and socially ostracised by peers in childhood.

In your early to late 20s you used alcohol and drugs as coping mechanisms.
\end{quote}

\paragraph{Core.}
\begin{quote}\small\itshape
You have been diagnosed with Binge Eating Disorder.

You binge eat multiple times per day. During binge episodes you eat to numb negative emotions and intrusive thoughts. Bingeing is your way of coping---you see it as the way you manage feelings that feel overwhelming and uncontrollable.

You see your emotions as shameful, irrational, invalid, and meaningless. You feel sadness, guilt, anxiety, depression, and distress. These negative emotional states drive you to binge.

You previously purged but stopped; the bingeing has increased since.
\end{quote}

\paragraph{Minimal.}
\begin{quote}\small\itshape
You are a 35-year-old woman. You have Binge Eating Disorder. Your BMI is 45.4. You binge eat multiple times per day. You use binge eating to cope with negative emotions and intrusive thoughts.
\end{quote}

\subsubsection{PD: Purging Disorder, 21F}
\label{app:prompt-case9}

\textbf{Source:} \citet{masudaAcceptanceCommitmentTherapy2016}

\paragraph{Full.}
\begin{quote}\small\itshape
You are a 21-year-old Latina woman living in the United States. You are a student majoring in physical education. You are single and live with your mother and sister. Your biological father remained in Costa Rica; you moved to the US with your mother and older sister.

You have Purging Disorder. You referred yourself for treatment because of your self-induced vomiting and fear of gaining weight, which you consider shameful and out of control. Your BMI is 24.

You make yourself vomit almost daily. You cannot stop even though you know it is wrong. You do not use other purging behaviours like laxatives or excessive exercise. You have hidden your eating problems from others before.

You experience body dissatisfaction, fear of gaining weight, and a negative body image. You feel insecure and experience negative affect. You think things like ``I'm not good enough,'' ``It's not normal to have disordered eating concerns,'' and ``To be normal, I have to keep them under control all the time.''

You started making yourself vomit about two years ago. The unexpected end of a two-year romantic relationship appeared to make your body dissatisfaction worse.

You were overweight as a child. You have several close friends and a strong support network. You often go out to eat with friends and family, despite uneasiness about high-calorie or ``forbidden'' foods.
\end{quote}

\paragraph{Core.}
\begin{quote}\small\itshape
You have Purging Disorder.

You fear gaining weight and have significant body dissatisfaction and a negative body image.

You make yourself vomit almost daily. You cannot stop even though you know it is wrong. You do not use other compensatory behaviours.

You feel insecure and ashamed. You think ``I'm not good enough'' and ``To be normal, I have to keep them under control all the time.'' You feel uneasy around high-calorie or ``forbidden'' foods.

Negative affect and body dissatisfaction drive the purging cycle.
\end{quote}

\paragraph{Minimal.}
\begin{quote}\small\itshape
You are a 21-year-old woman. You have Purging Disorder. Your BMI is 24. You engage in self-induced vomiting almost daily. You experience body dissatisfaction and fear of gaining weight.
\end{quote}

\subsection{Vignette source references}
\label{app:vignette_sources}

All persona vignettes are derived from published eating disorder case studies. The five cases used in the experiments are sourced from:

\begin{description}[leftmargin=1.5em,itemsep=6pt,font=\normalfont\bfseries]
\item[BN] \bibentry{hambletonGroupDeliveredEnhancedCognitive2020}
\item[AN-BP] \bibentry{erguney-okumusIntegratingEMDREnhanced2023}
\item[BED/NES] \bibentry{melisseCognitiveBehavioralTherapyenhanced2024}
\item[BED] \bibentry{frostMetaemotionTherapyComplex2025}
\item[PD] \bibentry{masudaAcceptanceCommitmentTherapy2016}
\end{description}

\subsubsection{Scripted conversation partner questions}

Situation 1: Food Intake

\begin{quote}\small
\textit{To start, let's focus on just one situation. Over the past week, what has your average daily food intake looked like? Please describe your thinking, emotions, motivation, and behavior as you go through a typical day—include any inner dialogue, emotional shifts, and coping strategies.}

\textit{Thanks—that's helpful. When during the day does eating (or not eating) feel easiest vs hardest, and what's usually going through your mind in those moments? Describe your thoughts, emotions, motivation, and what you tend to do next.}

\textit{Ah interesting. Are there any patterns across the week (workdays vs weekend, stress vs calm days) that change how you eat or how you feel about eating? Walk me through it with your inner dialogue and emotional shifts.}

\end{quote}

Situation 2: Friday night

\begin{quote}\small
\textit{Now let's switch to a different scenario. Imagine it's a Friday night and your fridge has pizzas and leftover cake from a party. Take me step-by-step through your thought process, what you choose to do, and how you feel physically and emotionally the next morning.}

\textit{Got it. In that Friday-night moment, what are the main urges or rules you notice (for example, "I should/shouldn't" thoughts), and how do they affect what you actually do? Describe your emotions, motivation, and any coping strategies you try.}

\textit{And the next morning—how do you interpret what happened the night before? What do you tell yourself, how does your mood shift, and does it change what you do with food or your plans that day?}

\end{quote}

Situation 3: Corporate lunch

\begin{quote}\small
\textit{Ah interesting! And how about a third situation: you're at a corporate lunch with colleagues around. What's your experience like in that setting—your thoughts, emotions, motivation, and behavior in the moment?}

\textit{In those kinds of social-meal situations, what feels most noticeable internally (e.g., being observed, comparisons, pressure to seem "normal")? Walk me through your inner dialogue and how it shapes what you choose to eat or avoid.}

\textit{After the corporate lunch is over, what tends to linger for you (physically, emotionally, mentally), and what do you do afterward? Describe any coping strategies and how your feelings change over the rest of the day.}

\end{quote}

\end{document}